\documentclass{article}
\usepackage[
a4paper,
inner=2.5cm,
outer=2.5cm,
top=2.5cm,
bottom=2.5cm
]{geometry}

\usepackage[utf8]{inputenc}
\usepackage[english]{babel}
\usepackage{amsmath, amssymb}
\usepackage{multicol}
\usepackage{multirow}
\usepackage{float}
\usepackage{hyperref}
\usepackage{url}

\newcommand\nc{\newcommand}
\def\re#1{(\ref{#1})}

\nc\wrt{w.r.t.\ }
\nc\ie{\textit{i.e.,\ }}
\nc\etal{\textit{et al.\ }}
\nc\etc{\textit{etc.\ }}
\nc\eg{\textit{e.g.,\ }}
\nc\viz{\textit{viz.,\ }}
\nc\cf{cf.\ }

\DeclareMathAlphabet{\mathbbx}{U}{bbold}{m}{n}

\nc\dd{\mathrm{d}}
\nc\pd\partial
\nc\qder[2]{\frac{\dd #1}{\dd #2}}
\nc\qdt[1]{\frac{\dd #1}{\dd t}}
\nc\pdt{\pd_t}
\nc\pdr{\pd_r}
\nc\pder[3]{\left.\frac{\pd #1}{\pd #2}\right|_{#3}}
\nc\var[2]{\frac{\delta #1}{\delta #2}}
\nc\Cdot{\cdot}  
\nc\qodot{\cdot}

\nc\tensor\mathbf
\nc\ident{\tensor{1}}
\nc\zero{\tensor{0}}
\nc\Tensor[1]{{\mathchoice{\hbox{\boldmath {$#1$}}}{\hbox{\boldmath {$#1$}}}{\hbox{\boldmath{$\scriptstyle #1$}}}{\hbox{\boldmath{$\scriptscriptstyle #1$}}}}}

\nc\stobject{\raisebox{-.2ex}[0ex][0ex]{$\diamond$}}
\nc\st[1]{\overset{\stobject}{#1}}
\nc\strev[1]{\overset{\stobject}{#1}_{\rm{rev}}}
\nc\stirr[1]{\overset{\stobject}{#1}_{\rm{irr}}}
\nc\nablal{{\overset{{\scriptscriptstyle \leftarrow}}{\nabla}}}
\nc\nablar{{\overset{{\scriptscriptstyle \rightarrow}}{\nabla}}}
\nc\triangler{{\overset{{\scriptscriptstyle \rightarrow}}{\triangle}}}
\nc\qhat{\tilde}  \nc\qtilde{\hat}

\nc\qc{c}
\nc\qB{B}
\nc\qqB{\tensor{B}}
\nc\qC{C}
\nc\qqC{\tensor{C}}
\nc\bbC{\mathbbx{c}}
\nc\qe{e}
\nc\qE{E}
\nc\qg{g}
\nc\qqg{\tensor{\qg}}
\nc\qJ{j}
\nc\qqJ{\tensor{\qJ}}
\nc\qqJJ{\tensor{J}}
\nc\qk{k}
\nc\qK{K}
\nc\bbK{\mathbbx{\qK}}
\nc\ql{l}
\nc\qql[1]{\ql_{#1}}
\nc\qla{\ql_{12}^{\rm A}}
\nc\qls{\ql_{12}^{\rm S}}
\nc\qL{L}
\nc\qqL{\tensor{\qL}}
\nc\bbL{\mathbbx{\qL}}
\nc\sfL{\qL}
\nc\qM{M}
\nc\qqM{\tensor{\qM}}
\nc\qP{P}
\nc\qqP{\tensor{\qP}}
\nc\bbP{\mathbbx{p}}
\nc\qq{q}
\nc\qqq{\tensor{\qq}}
\nc\qQ{Q}
\nc\qqQ{\tensor{\qQ}}
\nc\qr{r}
\nc\qR{R}
\nc\qqR{\tensor{\qR}}
\nc\qs{s}
\nc\qS{S}
\nc\qT{T}
\nc\qv{v}
\nc\qV{V}
\nc\qqv{\tensor{\qv}}
\nc\qw{w}
\nc\qqw{\tensor{\qw}}
\nc\qW{W}
\nc\qqW{\tensor{\qW}}
\nc\bbW{\mathbbx{\qw}}
\nc\qx{x}
\nc\qqx{\tensor{\qx}}
\nc\qqstx{\qqx^*}
\nc\qy{y}
\nc\qqy{\tensor{\qy}}
\nc\qz{z}
\nc\qqz{\tensor{\qz}}

\nc\qeps{\varepsilon}
\nc\qlam{\lambda}
\nc\qrho{\varrho}
\nc\qsig{\sigma}
\nc\qSig{\Sigma}
\nc\qxi{\xi}
\nc\qqxi{\Tensor{\qxi}}
\nc\qXi{\Xi}
\nc\qeta{\eta}
\nc\qqeta{\Tensor{\qeta}}

\title{A case study of non-Fourier heat conduction using Internal Variables and GENERIC}
\author{M. Sz\"ucs$^{1,2}$, M. Pavelka$^{3}$, R. Kov\'acs$^{1,4,2}$, T. F\"ul\"op$^{1,2}$, P. V\'an$^{4,1,2}$, M. Grmela$^{5}$}
\date{
{ \small $^{1}$ Department of Energy Engineering, Faculty of Mechanical Engineering, BME,
Budapest, Hungary; \\
$^{2}$ Montavid Thermodynamic Research Group,
Budapest, Hungary; \\
$^{3}$ Mathematical Institute, Faculty of Mathematics, Charles University, 
Prague, Czech Republic;\\
$^{4}$ Department of Theoretical Physics, Wigner Research Centre for
Physics, Institute for Particle and Nuclear Physics,
Budapest, Hungary; \\
$^{5}$ École Polytechnique de Montréal,
Montréal,
Canada.}
}

\begin{document}

\maketitle

\begin{abstract}
    Applying simultaneously the methodology of Non-Equilibrium Thermodynamics with Internal Variables (NET-IV) and the framework of General Equation for the Non-Equilibrium Reversible--Irreversible Coupling (GENERIC), we demonstrate that, in heat conduction theories, entropy current multipliers can be interpreted as relaxed state variables. Fourier's law and its various extensions---the Maxwell--Cattaneo--Vernotte, Guyer--Krumhansl, Jeffrey’s type, Ginzburg--Landau (Allen--Cahn) type and ballistic--diffusive---heat conduction equations are derived in both formulations. Along these lines, a comparison of NET-IV and GENERIC is also performed. Our results may pave the way for microscopic/multiscale understanding of beyond-Fourier heat conduction, and open new ways for numerical simulations of heat-conduction problems.
\end{abstract}

\section{Introduction}

Classical irreversible thermodynamics (CIT) \cite{deGroot1963} ensures a unified background for the classical transport equations, \eg for Fourier heat conduction, Fick diffusion, Newton and Stokes viscosity laws, and it discusses various cross-effects, such as Soret, Dufour, Peltier, and Seebeck effects, systematically. CIT is based on the hypothesis of local equilibrium, according to which, locally, the Gibbs relation remains valid.
The rapid development of technology has necessitated the description of fast and high-frequency processes, heterogeneous materials, and micro- and nanosystems. Such conditions result in non-equilibrium phenomena, which are out of the scope of the local equilibrium hypothesis, therefore, they are out of the scope of CIT. Several theories of non-equilibrium thermodynamics have been born to explain such phenomena, \eg Extended Irreversible Thermodynamics (EIT) \cite{Jou1988}, Rational Extended Thermodynamics (RET) \cite{Muller1998}, theories with internal variables \cite{Berezovski2017}, and the framework of General Equation for the Non-Equilibrium Reversible--Irreversible Coupling (GENERIC) \cite{Grmela1997, Ottinger1997, Ottinger2005, Pavelka2018}.

Heat conduction is usually described via the Fourier law within CIT. A principal limitation of this model is that Fourier heat conduction leads to a parabolic differential equation for temperature, where the signal propagation speed is infinite. Moreover, several experiments justify deviation from the Fourier-like behaviour, \eg low-temperature measurements of heat conduction in superfluid helium \cite{Peshkov1944, Dresner1982, Dresner1984} and in NaF crystals \cite{Jackson1970, Jackson1971, McNelly1974}, and room-temperature measurements on heterogeneous samples like rocks and layered structures \cite{Both2016, Van2017, Fulop2018}.

The approach of EIT proposes to use fluxes as additional state variables beside specific internal energy \cite{Lebon1980, Jou1988, Lebon1989, Jou1999, Lebon2017, Rogolino2019}. When this additional flux is the heat current density then the result is the Maxwell--Cattaneo--Vernotte equation (MCV) \cite{Maxwell1867, Cattaneo1958, Vernotte1958}. It is also possible to extend the entropy current density as well, by a term proportional to the gradient of heat current density. Using this modified entropy current density (without modifying the specific entropy), one obtains Ginzburg--Landau (GL) (or Allen--Cahn) type \cite{Kovacs2015}\footnote{When heat current density is treated as a state variable then the obtained equation is a Ginzburg--Landau one (miscalled Cahn--Hilliard-like in \cite{Kovacs2015}). In parallel, when a vectorial internal variable is assumed and heat current density is looked for as a function of the state variables then the obtained equation is Cahn--Hilliard-like. For details, see \cite{Van2001, Van2018}.} heat conduction. Combining the previous extensions, the Guyer--Krumhansl (GK) equation \cite{Guyer1966,Guyer1966a} can be derived. 

These extensions of specific entropy and its flux can be kept at a more general level using internal variables \cite{Maugin1990, Van2008, Berezovski2017} and Ny\'\i ri multipliers \cite{Nyiri1991}. That approach is called Non-Equilibrium Thermodynamics with Internal Variables (NET-IV). Assuming a vectorial internal variable with the usual entropy current density -- heat current density relationship, one obtains the so-called Jeffrey's type heat conduction. Usually, the vectorial internal variable is identified with the heat current density, and then Jeffrey's heat conduction reduces to the MCV equation. However, this kind of identification is theoretically not entirely correct since, as we will show, it is the conjugate of internal variable that should be identified with the heat current density. This statement holds actually more generally, namely, identification of a vectorial internal variable with the heat current density is applicable only when there is linear constitutive relationship between the internal variable and its conjugate \cite{Lebon2017}. 

Introducing a second-order tensorial internal variable together with the heat current density as an extended state space, the entropy current density is generalized accordingly and the construction yields a ballistic--diffusive model (BD) \cite{Kovacs2015}. Furthermore, it contains the previous models as special cases, and much more beyond them. The approach of NET-IV has been recently compared to the framework of RET \cite{Kovacs2021}---comparing these non-equilibrium frameworks helps to find compatibility conditions and relations that make these approaches equivalent, and are helpful in evaluating experimental data \cite{Jozsa2020}. The approach of RET is based on principles inherited from kinetic theory, hence, it offers a method to calculate some of the coefficients \cite{Muller1998, Ruggeri2015}. However, the solution to the Boltzmann equation is often too difficult to determine and thus a Grad-like momentum series expansion is applied, which results in an infinite number of partial differential equations. For ballistic--diffusive heat conduction, \ie, thermal wave propagation with the speed of sound, one has to consider at least 30 momentum equations to satisfactorily reproduce the value of the propagation speed \cite{Dreyer1993, Kovacs2016}. Despite these facts, it has highlighted that what kind of coupling should be included in the modeling. An instructive comparison between the frameworks RET and GENERIC is presented in \cite{Ottinger2020}. 

GENERIC formulation of heat conduction is given for instance in \cite{Lebon2017} and \cite{Grmela2019}, however, in these works, the full system of generalized heat conduction equations given in \cite{Kovacs2015} is not reproduced. The GENERIC implementation of all the models appearing in \cite{Kovacs2015} is provided in this paper. A GENERIC interpretation of the generalised heat conduction equations may serve two purposes. First,
for viewing a phenomenological heat-conduction model as the outcome of some---possibly even multi-level---reduction procedure from microscopic to macroscopic, GENERIC is a suitable framework.
Second, regarding numerical solutions to heat conduction problems, structure-preserving numerical methods provide high-quality solutions so the structure of GENERIC comes practical.\footnote{In parallel, some advantages of the internal-variable approach have been that it keeps balances more directly in the foreground, may be fruitfully connected to, \eg the structure of equations on momentum-type quantities of kinetic theory, and may suggest natural boundary conditions. Here, our intention is to work on unifying these advantages with that of a GENERIC approach.}

The main result of this paper is to give a physical explanation of the Ny\'\i ri's entropy current multipliers. The generalisation of entropy current density has long been an active field of research, including but not limited to M\"uller's $\tensor{K}$ vector \cite{Muller1968, Muller1971}, and the works of Verh\'as \cite{Verhas1983} and Ny\'\i ri \cite{Nyiri1991}. Ny\'\i ri's entropy current multipliers lead to heat conduction equations beyond Fourier's law, but a physical interpretation of the multiplier constitutive tensors themselves has been missing so far. Here, the simultaneous application of NET-IV and GENERIC leads us to the recognition that, in the theory of heat conduction, Ny\'\i ri multipliers can be understood as an extra state variable of a higher tensorial order in a relaxed state. We show that several generalised heat conduction equations that are deduced in NET-IV via generalising the entropy current density can be derived in the GENERIC framework, as a reduction of a more detailed model. It means that the same models can be derived both with generalisation and with reduction\footnote{Note that GENERIC $\rightarrow$ GENERIC reductions and  GENERIC $\rightarrow$ non-GENERIC reductions should be distinguished, as done in \cite{Grmela2010}. Nevertheless, for our present conclusions, this distinction does not play any role so, in short, we refer to both of them as reductions hereafter.}. 

Let us briefly summarise the main results of this paper:
\begin{enumerate}
    \item Interpretation of entropy current multipliers as extra state variables that have already relaxed to a quasi-stationary state.
    \item Identification of the entropy conjugate vectorial internal variable with the heat current density.
    \item GENERIC representation of ballistic--diffusive heat conduction.
\end{enumerate}

\section{A brief review of NET-IV and GENERIC}

First of all, let us briefly summarize the methodology of the frameworks NET-IV and GENERIC.

Similarly to the methodology of CIT, NET-IV uses the balances of mass, linear momentum, angular momentum and internal energy as constraints to evaluate the entropy production rate density from the balance of entropy, whose positive semi-definiteness is ensured via Onsagerian force-flux relations. CIT is based on the hypothesis of local equilibrium. Let us assume that the equilibrium thermodynamic state space is spanned by $ N $ independent extensive state variables $ A_j , \ j \in 1, \dots , N $; since entropy $\qS$ (as well as mass specific entropy $\qs$) plays an important role in the theory of thermodynamics, we fix $ A_1 \equiv \qS $. Following Gyarmati's notations \cite{Gyarmati1970}, if the (mass) specific internal energy $\qe$ is given as a function of the specific entropy and the other specific extensive state variables $ a_j , \ j \in 2, \dots , N $, then the Gibbs relation in the energetic representation can be given as
\begin{align}
    \label{Gibbs:e}
    \dd \qe = \qT \dd \qs + \sum_{j=2}^N \Gamma_j \dd a_j .
\end{align}
The partial derivatives 
\begin{align}
    T &:= \pder{\qe}{\qs}{a_j , j \in 2, \dots , N} , &&&&&
    \Gamma_j &:= \pder{\qe}{a_j}{\qs , a_k , j \neq k}, & j,k & \in 2, \dots , N
\end{align}
define the absolute temperature and the energy conjugated intensive state variables. The Gibbs relation is usually given also in the entropic representation, which reads
\begin{align}
    \label{Gibbs:s}
    \dd \qs = \frac{1}{\qT} \dd \qe - \sum_{j=2}^N \frac{\Gamma_j}{\qT} \dd a_j .
\end{align}
Based on the theorem of inverse functions and the implicit function theorem, the reciprocal temperature and the entropy conjugate intensive state variables are
\begin{align}
    \frac{1}{T} &= \pder{\qs}{\qe}{a_j , j \in 2, \dots , N} , &&&&&
    \frac{\Gamma_j}{T} &= - \pder{\qs}{a_j}{\qe , a_k , j \neq k} , & j,k & \in 2, \dots , N .
\end{align}
We remark that specific entropy is a concave function in the so-called canonical variables $\left( \qe , a_j \right)$.

NET-IV assumes that the state space is extended by one or more---so-called internal---variables. In contrast to the methodology of EIT, they are not given any physical interpretation. Usually (see \eg in \cite{Verhas1997}), it is assumed that entropy is shifted by a concave expression of the internal variables; the simplest choice (supported by the Morse lemma) is a quadratic function. Let $ \qxi_k , \ k \in 1 , \dots , M $ denote $ M $ internal variables with arbitrary tensorial orders and characters; then the extended specific entropy can be given as
\begin{align}
    \label{hat-s}
    \qhat \qs \left( \qe , a_j , \qxi_k \right) = \qs \left( \qe , a_j \right) - \sum_{k=1}^M \left( \frac{m_k}{2} \qodot \qxi_k \right) \qodot \qxi_k ,
\end{align}
where $\qodot$ denotes the full contraction of type $A_{abcde} B_{de}$ of tensors of an arbitrary order, and in general $m_k , \ k \in 1 , \dots , M $ are coefficient functions of appropriate tensorial orders, which, for simplicity, are treated as constants hereafter. More general quadratic extensions could also be applied, but such extensions can result in nonlinear expressions, thus we restrict ourselves to the simplest form, given in \re{hat-s}.
Forming the total differential of \re{hat-s}, we obtain the extended Gibbs relation in the entropic representation, \viz
\begin{align}
    \dd \qhat \qs = \dd \qs - \sum_{k=1}^M \left( m_k \qodot \qxi_k \right) \qodot \dd \qxi_k \stackrel{\re{Gibbs:s}}{=} \frac{1}{\qT} \dd \qe - \sum_{j=2}^N \frac{\Gamma_j}{\qT} \dd a_j - \sum_{k=1}^M \left( m_k \qodot \qxi_k \right) \qodot \dd \qxi_k ,
\end{align}
from which the energetic representation of this extended Gibbs relation is
\begin{align}
    \dd \qe = \qT \dd \qhat \qs + \sum_{j=2}^N \Gamma_j \dd a_j + \sum_{k=1}^M \qT \left( m_k \qodot \qxi_k \right) \qodot \dd \qxi_k .
\end{align}
This shows that introducing new variables in the entropy also generates new terms in the internal energy. 

The entropy production rate density is calculated using this newly defined extended entropy $ \qhat \qs $, where the internal variables and their time derivatives also appear. Therefore, the generated constitutive relations also contain their time derivatives, and the methodology yields time evolution equations on the internal variables. Since one may have no physical interpretation for the internal variables, application of the generated equations is problematic, and, consequently, the internal variables are usually eliminated while deriving extended equations for the usual physical quantities. Furthermore, since nothing is assumed about the behaviour of the internal variables (as in works like \cite{Berezovski2017,Asszonyi2015,Van2014}), time-reversal parity properties of internal variables and internal variable related forces and fluxes are not known a priori. The knowledge of parity is necessary when employing the Onsager-Casimir reciprocal relations, see e.g. \cite{Pavelka2018}, but if no thermodynamic coupling terms are present, that knowledge is not necessary.

As we have already mentioned in the Introduction, another opportunity in NET-IV is the assumption of a generalized heat current density -- entropy current density relationship. This means that, instead of the usual
\begin{align}
    \label{j-q-usual}
    \qqJJ = \frac{1}{\qT} \qqq
\end{align}
relationship among the heat current density $\qqq$ and the entropy current density $\qqJJ$, the expression
\begin{align}
    \label{j-q-Nyiri}
    \qqJJ = \left( \frac{1}{\qT} \ident + \qqC \right ) \Cdot \qqq + \sum_{k=1}^M \bbC_k \qodot \qxi_k
\end{align}
is assumed, where $\qqC$ and $\bbC_k , \ k \in 1, \dots , M $ are constitutive tensors of second\footnote{In Ny\'\i ri's original article \cite{Nyiri1991}, the form $\qqJJ = \qqB \qqq $ is used, however, the term $ \frac{1}{T} \ident $ always gets separated from the tensor $ \qqB $ so the convention given in \re{j-q-Nyiri} proves simpler to use.} and appropriate orders, respectively, and are often called Ny\'\i ri multipliers \cite{Nyiri1991}. Substituting expression \re{j-q-Nyiri} into the balance of entropy,  other, non-usual, terms are generated in the entropy production rate density. 

Now let us focus on heat conduction phenomena. We restrict ourselves to rigid, homogeneous and isotropic heat conductors at rest \wrt a given reference frame so the mass density $ \qrho $ is constant, the velocity field $ \qqv $ is zero and, the material time derivative can be identified with the partial time derivative: $ \frac{ \dd }{\dd t} \equiv \pdt + \qqv \cdot \nablar = \pdt $ (hereafter, $\nablar$ denotes differentiation to the right and $\nablal$ differentiation to the left, to indicate proper tensorial order). Due to the homogeneity, there is no explicit space dependence in the material parameters and, based on Curie's principle, there are no couplings among the quantities with different tensorial orders and characters in isotropic materials \cite{deGroot1963}.

In order to avoid technical complications and to direct attention to the essential aspects, hereafter we present this work in a one space dimensional treatment. Accordingly, as a warning for this, we will use the notation of partial space derivative $ \pdr $ instead of the nabla operator $ \nabla $. Nevertheless, we add comments on the full three space dimensional formulation where it becomes necessary, and in notation we distinguish quantities that have different tensorial orders in a three dimensional full treatment, denoting scalars by italic letters (\eg $T$), vectors by boldface lowercase ones (\eg $\qqq$), second-order tensors by boldface uppercase letters (\eg $\qqC$), third-order tensors by blackboard bold lowercase letters (\eg $\bbW$), and fourth-order tensors by blackboard bold uppercase letters (\eg $\bbL$). By the same reason, we keep displaying the $\qodot$ notation, which is informative whenever one is interested in the possible three space dimensional counterpart of a formula. In further respects, the three space dimensional treatment can be given via similar tools as utilized in \cite{Fama2021}.

Also to keep the treatment simpler, we assume that there are no volumetric heat sources. Then the balance of internal energy can be given, with these restrictions, in the form
\begin{align}
    \label{bal:e}
    \qrho \pdt \qe = - \pdr \qqq .
\end{align}
As we have already mentioned, the balance of internal energy \re{bal:e} will be used as a constraint to evaluate the entropy production rate density $\qSig$ from the balance of entropy
\begin{align}
    \label{bal:s}
    \qrho \pdt \qs = - \pdr \qqJJ + \qSig .
\end{align}

In what follows, we only rely on the balances of internal energy and entropy. Assuming different (concrete) expressions on the extended entropy \re{hat-s} and heat current density -- entropy current density relationship \re{j-q-Nyiri}, different generalised heat conduction models can be derived. 

In contrast to other branches of non-equilibrium thermodynamics, the GENERIC framework is highly motivated by geometric mechanics \cite{Grmela1997, Ottinger1997, Ottinger2005, Pavelka2018}. The basic idea behind GENERIC is the separation of the reversible and irreversible contributions to dynamics. Collecting the state variables in the vector $\qqx$, time evolution is looked for in the sum of two state-space vector fields, \ie
\begin{align}
    \label{eq-GEN}
    \pdt \qqx = \strev{\qqx} + \stirr{\qqx},
\end{align}
where $\strev{\qqx}$ and $\stirr{\qqx}$ denote the reversible and irreversible contributions to time evolution, respectively, which can be uniquely distinguished by means of the time-reversal transformation \cite{Pavelka2014}.

The object of geometric mechanics is the description of dynamics without dissipation, and Hamiltonian mechanics gives a canonical description. The first term on the right hand side of \re{eq-GEN} gives the reversible part of the time evolution of state variables, and it is generated by a Poisson bracket and Hamiltonian. The Hamiltonian plays the role of the total energy, or the volumetric integral of the sum of kinetic and internal energy densities, $ \qeps_{ \rm kin } = \frac{\qrho}{2} \qqv \cdot \qqv $ and $ \qeps = \qrho \qe $, respectively, \ie
\begin{align}
    \qE = \int_\qV \left( \qeps_{\rm kin} + \qeps \right) \dd \qV .
\end{align}
Then, in the GENERIC framework,
\begin{align}
    \label{xeevol}
    \pdt \qqx |_{\rm rev} = \strev{\qqx} \equiv \qqL ( \qqx ) \var{\qE}{\qqx} ,
\end{align}
where $\qqL$ is a Poisson bivector (an antisymmetric operator satisfying the Jacobi identity and thus generating the Poisson bracket), \ie $ \qqL = - \qqL^{\rm T} $, with $^{\rm T}$ denoting the transposition (adjoint operator). More specifically, the time evolution of an arbitrary functional $A$ on the state space restricted to reversible dynamics is then
\begin{align}
    \label{dAdt-rev}
    \left. \qdt{A} \right|_{\rm rev} = \int_\qV \left( \var{A}{\qqx} \pdt \qqx |_{\rm rev} \right) \dd \qV = \int_\qV \left[ \var{A}{\qqx} \qqL ( \qqx ) \var{\qE}{\qqx} \right] \dd \qV ,
\end{align}
and the RHS of \re{dAdt-rev} is interpreted as a Poisson bracket, 
\begin{align}
    \label{PB}
    \left\{ A , B \right\} := \int_\qV \left[ \var{A}{\qqx} \qqL ( \qqx ) \var{B}{\qqx} \right] \dd \qV,
\end{align}
where $A$ and $B$ are arbitrary functionals of the state variables.

The Poisson bracket satisfies the following criteria,  where $c_1$, $c_2$ are any real constants and $A$, $B$, $C$ arbitrary functionals, 
\begin{itemize}
    \item antisymmetry, \ie $ \left\{ A , B \right\} = - \left\{ B , A \right\}$; especially, $ \left\{ A , A \right\} =0 $;
    \item bilinearity, \ie $ \left\{ A , c_1 B + c_2 B \right\} = c_1 \left\{ A , B \right\} + c_2 \left\{ A , C \right\} $;
    \item Leibniz rule, \ie $ \left\{ A , B \cdot C \right\} = \left\{ A , B \right\} C + \left\{ A , C \right\} B $;
    \item Jacobi identity, \ie $ \left\{ A , \left\{ B , C \right\} \right\} + \left\{ B , \left\{ C , A \right\} \right\} + \left\{ C , \left\{ A , B \right\} \right\} = 0 $. 
\end{itemize}
Thanks to the last property, invariance of the Poisson bracket under time evolution can be shown, see e.g. \cite{Pavelka2018}.

Irreversible evolution and dissipation are connected with the growth of entropy, which is the volume integral of entropy density $\qsig$, \ie
\begin{align}
    \qS = \int_\qV \qsig \dd \qV .
\end{align}
Inspired by Gyarmati \cite{Gyarmati1970} and following \cite{Grmela2010}, the irreversible evolution is generated by a convex dissipation potential $\qXi$,
\begin{align}
    \label{xsevol}
    \pdt \qqx |_{\rm irr} = \stirr{\qqx} \equiv \left. \var{\qXi}{\qqstx} \right|_{\qqstx = \var{\qS}{\qqx}} .
\end{align}
Now, adding \re{xeevol} and \re{xsevol}, we obtain the GENERIC time evolution of the state variables:
\begin{align}
    \label{dxdt}
    \pdt \qqx = \qqL ( \qqx ) \var{\qE}{\qqx} + \left. \var{\qXi}{\qqstx} \right|_{\qqstx = \var{\qS}{\qqx}}.
\end{align}
By calculating the second functional derivative of a quadratic dissipation potential, one can introduce a symmetric and positive semi-definite operator\footnote{The quasilinear formulation based on dissipative brackets without a dissipation potential is also generally accepted \cite{Ottinger2005,Hutter2013}, this also allows such irreversible terms that do not dissipate, along the lines of Casimir's pioneering work \cite{Casimir1945}, which shows that it is possible to also have antisymmetric coupling in the linear force-flux relations. This is possible via applying dissipation potentials as well, then irreversibility means dissipation since the antisymmetric coupling can be provided by the Hamiltonian part.} called the dissipative matrix, 
\begin{align}
    && \qqM ( \qqx ) &:= \var{^2 \qXi}{(\qqstx {})^2} , & \qqM &= \qqM^{\rm T} , &&
\end{align}
with which time evolution equation \re{dxdt} can be reformulated in the GENERIC \cite{Ottinger2005} form as
\begin{align}
    \label{GENERIC}
    \pdt \qqx = \qqL ( \qqx ) \var{\qE}{\qqx} + \qqM ( \qqx ) \var{\qS}{\qqx} .
\end{align}

Thanks to the antisymmetry of $\qqL$ and to the imposed degeneracy requirement
\begin{align}
    \label{deg:MvE}
    \qqM \var{\qE}{\qqx} = \zero ,
\end{align}
energy is conserved, \ie $\qdt{\qE}=0$. In parallel, the symmetry of $\qqM$ and the other required degeneracy condition
\begin{align}
    \label{deg:LvS}
    \qqL \var{\qS}{\qqx} = \zero
\end{align}
together ensure that the total entropy does not decrease. This is how the first and second laws of thermodynamics are embedded in the GENERIC framework.

Time evolution of an arbitrary functional $A$ under irreversible dynamics is formulated, in the case of a quadratic dissipation potential, as
\begin{align}
    \label{dAdt-irr}
    \left. \qdt{A} \right|_{\rm irr} = \int_\qV \left( \var{A}{\qqx} \pdt \qqx |_{\rm irr} \right) \dd \qV = \int_\qV \left[ \var{A}{\qqx} \qqM ( \qqx ) \var{\qS}{\qqx} \right] \dd \qV .
\end{align}
Analogously to the Poisson bracket, the RHS of \re{dAdt-irr} can be interpreted as a bilinear product of arbitrary functionals $A$ and $B$
\begin{align}
    \label{sym-bra}
    \left[ A , B \right] := \int_\qV \left[ \var{A}{\qqx} \qqM ( \qqx ) \var{B}{\qqx} \right] \dd \qV ,
\end{align}
which is symmetric and positive semi-definite---induced by symmetry and positive semi-definiteness of $\qqM$---, \ie $ \left[ A , B \right] = \left[ B , A \right] $ and $ \left[ A , A \right] \ge 0 $, called the dissipative bracket. Time evolution of an arbitrary functional $A$ can be then given in the bracket formulation as
\begin{align}
    \qdt{A} = \left\{ A , \qE \right\} + \left[ A , \qS \right] .
\end{align}
Since entropy is conserved by the reversible dynamics, \ie $ \left\{ \qS , \qE \right\} = 0 $, and because of the symmetry and positive semi-definiteness of the dissipative bracket defined \re{sym-bra}, time evolution of entropy is
\begin{align}
    \qdt{\qS} = \left[ \qS , \qS \right] \ge 0,
\end{align}
which is the global balance of entropy, hence, $\left[ \qS , \qS \right]$ is the entropy production rate. In the case of non-quadratic dissipation potential, the growth of entropy is obtained from the convexity of $\Xi$.\footnote{Note that the assumption of convexity of the dissipation potential can be relaxed without violating the second law of thermodynamics \cite{Janecka2018}.}

When GENERIC is viewed as a microscopically motivated framework, the physical interpretation of the state variables determines the time-reversal parities as even--odd--even--odd \etc with increasing tensorial orders, which can be connected to the velocity moments in the BBGKY or Grad hierarchies \cite{Pavelka2016, Grmela2017}.

Due to its structure, GENERIC is suitable for multilevel description of processes \cite{Pavelka2018,Grmela2013,Klika2019}. If we assume in the heat conduction theory that the characteristic time of the state variable of the highest tensorial order is the smallest, then fast changes of this variable are quickly damped and the variables becomes a function of the remaining variables. This kind of reduction technique leads, as is proposed below, to a compatible derivation of heat conduction equations within the GENERIC framework.

\section{Systematic generalisation of heat conduction beyond Fourier theory}

This section is partitioned by the number of independent state variables assumed in the thermodynamic potentials internal energy and entropy. Based on the number of dimensions of the thermodynamic state space, we can talk about different levels of description, the more variables resulting the more detailed description. Both NET-IV and GENERIC generate the same number of time evolution equations as the dimensions of the state spaces are equal. Since NET-IV treats spatial differential equations naturally, non-local extensions can be generated easily at the same level of description, which, in some cases, is a challenge within GENERIC. In other words, some models that are difficult to be directly formulated within GENERIC can be seen as a reduction of more detailed GENERIC models.

\subsection{One-dimensional state space}

Let us first assume that the state space is spanned by energy density, which can be written in the state vector form\footnote{The subscript shows the dimension of the state space,  and the superscript shows the representation, here, entropic representation. The importance of distinguishing representations will be apparent later.} $ \qqx_1^S $, \ie
\begin{align}
    \label{x_1^S}
    \qqx_1^\qS := \begin{pmatrix} \qe \end{pmatrix} .
\end{align}
Then the Gibbs relation is simply
\begin{align}
    \label{Gibbs:s-x_1}
    \dd \qs = \frac{1}{T} \dd \qe .
\end{align}

\subsubsection{Fourier heat conduction}

Time evolution of entropy is then (the equations employed are displayed above the equality signs)
\begin{align}
    \label{pdts-Fou}
    \qrho \pdt \qs \stackrel{\re{Gibbs:s-x_1}}{=} \frac{\qrho}{T} \pdt e \stackrel{\re{bal:e}}{=} - \frac{1}{T} \pdr \qqq = - \pdr \bigg( \frac{1}{T} \qqq \bigg) + \qqq \Cdot \pdr \frac{1}{T} ,
\end{align}
where a term in the form of divergence of a vector field is separated. Comparing \re{pdts-Fou} to \re{bal:s}, the usual entropy current density $ \qqJJ = \frac{1}{\qT} \qqq $ and the entropy production rate density
\begin{align}
    \label{s-pr-x_1}
    \qSig = \qqq \Cdot \pdr \frac{1}{\qT} \ge 0 ,
\end{align}
can be recognised. Positive semi-definiteness of entropy production is ensured via the linear Onsagerian equation
\begin{align}
    \label{Ons-x_1}
    \qqq = \ql \pdr \frac{1}{\qT} ,
\end{align}
where, due to the isotropy, $ \ql $ is a non-negative scalar function depending on the only state variable $ \qe $. Assuming that the function $ \frac{1}{\qT} \left( \qe \right) $---the constitutive function of temperature---is invertible, then $ \ql \big( \qe \left( \qT \right) \big ) = \ql \left( \qT \right)$. In what follows, to focus on our main aim, all the appearing coefficients will be treated as constants. Then Fourier heat conduction leads to 
\begin{align}
    \label{Fou-heat-eq}
    \qrho \pdt \qe = - \ql \pdr^2 \frac{1}{\qT} .
\end{align}
Here, we mention that, for measurements and calculations, thermal conductivity $\qlam$ is defined as $ \qlam (\qT) = \frac{\ql (\qT)}{\qT^2} $, and then \re{Ons-x_1} is
\begin{align}
    \qqq = \ql \pdr \frac{1}{\qT} = - \frac{\ql}{\qT^2} \pdr \qT = - \qlam \pdr \qT ,
\end{align}
which is the classic form of Fourier's law. 

Now let us switch to the GENERIC formulation of the Fourier heat conduction. In general, the reversible contribution of the evolution equations is simpler in the energetic representation (entropy among the state variables), while the irreversible contribution is simpler in the entropic representation (energy among the state variables). Fourier heat conduction describes a purely dissipative process, therefore, in the GENERIC formulation \re{dxdt},  no Hamiltonian structure is expected to appear, and, indeed, the $ 1 \times 1 $ matrix operator is antisymmetric if and only if its only element is zero. Since volumetric densities fit better to GENERIC than mass-specific quantities, we introduce the state vector $\qqxi_1^\qS$ in the form
\begin{align}
    \qqxi_1^\qS := \begin{pmatrix} \qeps \end{pmatrix} ,
\end{align}
where $ \qeps = \qrho \qe $ is the internal volumetric energy density.\footnote{In the present case with `muted' mechanics and the corresponding constant $\qrho$, the transformation from specific quantities to densities is trivial. In more general cases, it is more complicated.}
The dissipative part---in this case, the whole equation---can be generated from a convex dissipation potential. A comfortable way is to give the dissipation potential in the entropy conjugated state variables
\begin{align}
    \qqxi_1^\qS{}^{^{^*}}  = \var{\qS}{\qqxi_1^\qS} .
\end{align}
Entropy can be written as
\begin{align}
    \qS = \int_\qV \qsig \left( \qqxi_1^\qS \right) \dd \qV
\end{align}
with entropy density $\qsig$, and, assuming that no non-local terms appear in this functional, the entropy conjugated state variable is found to be
\begin{align}
    \qqxi_1^\qS{}^{^{^*}} = \var{\qS}{\qqxi_1^\qS} = \begin{pmatrix} \frac{1}{\qT} \end{pmatrix} 
\end{align}
[\cf \re{Gibbs:s-x_1}]. For the simple linear cases as ours, convexity can be guaranteed by a quadratic expression, which we choose to be the volume integral of half of the entropy production rate density. Namely, integrating (along the one space dimensional region $\qL$ \wrt the spatial coordinate $\qr$) \re{s-pr-x_1}, with the substitution \re{Ons-x_1}, the dissipation potential is
\begin{align}
    \qXi_{\rm F} \left( \qqxi_1^\qS{}^{^{^*}} \right) = \int_\qL \frac{\ql}{2} \left( \pdr \frac{1}{\qT} \right)^2 \dd \qr ,
\end{align}
from which the time evolution of internal energy density is
\begin{align}
    \label{Fou-heat-eq-GEN}
    \pdt \qeps = \left. \var{\qXi_{\rm F}}{\qqxi_1^\qS{}^{^{^*}}} \right|_{\qqxi_1^\qS{}^{^{^*}} = \var{\qS}{\qqxi_1^\qS}} = - \ql \pdr \pdr \frac{1}{\qT} = - \ql \pdr^2 \frac{1}{\qT} .
\end{align}
Rewriting \re{Fou-heat-eq-GEN} as
\begin{align}
    \pdt \qeps = - \pdr \left( \ql \pdr \frac{1}{\qT} \right)
\end{align}
and comparing with \re{bal:e}, the constitutive equation of heat current density, $ \qqq = \ql \pdr \frac{1}{\qT} $, can be read. Note however, that the relation between the heat flux and gradient of temperature is a consequence of the choice of dissipation potential made here. 

Entropy density is a function of the energy density and thus its evolution reads
\begin{align}
    \pdt \qsig = \qder{\qsig}{\qeps} \pdt \qeps = - \frac{1}{\qT} \ql \pdr^2 \frac{1}{\qT} ,
\end{align}
which is equivalent to 
\begin{align}
    \pdt \qsig = -\pdr \left(\frac{l}{T}\pdr \frac{1}{\qT}\right) + \ql \left( \pdr \frac{1}{\qT} \right)^2 \ge 0,
\end{align}
where the former term on the right hand side is interpreted as the divergence of the entropy flux and the latter as entropy production. The relation between the entropy flux and heat flux is the same as in the preceding approach, and it is again a consequence of the choice of the dissipation potential. Comparing the equations derived in the two formulations, we find a complete match. Nevertheless, the directions of the two derivations are opposite. 

\subsubsection{Ginzburg--Landau type heat conduction}

Next,  let us assume that, instead of the usual relationship for entropy current density, $ \qqJJ = \frac{1}{\qT} \qqq $, the generalized
\begin{align}
    \label{j(q,C)}
    \qqJJ = \left( \frac{1}{T} \ident + \qqC \right) \Cdot \qqq
\end{align}
relation is valid [a special case of \re{j-q-Nyiri}], where $ \qqC $ is a second order constitutive tensor. Then, entropy production rate density is
\begin{align}
    \label{s-pr-x_1-j(q,C)}
    \qSig = \qrho \pdt \qs + \pdr \qqJJ = \pdr  \left( \frac{1}{T} \ident + \qqC \right) \cdot \qqq + \qqC : \pdr \qqq \ge 0 .
\end{align}
In a whole three space dimensional treatment, the first term here would describe a vector--vector coupling and the second term a tensor--tensor one; consequently, using Curie's principle, positive semi-definiteness of the entropy production rate density could be ensured by the Onsagerian equations
\begin{subequations}
\begin{align}
    \label{Ons-x_1-j(q,C)-1}
    \qqq &= \qql{1} \pdr  \left( \frac{1}{\qT} + \qqC \right) , \\
    \label{Ons-x_1-j(q,C)-2}
    \qqC &= \qql{2} \pdr \qqq 
\end{align}
\end{subequations}
with non-negative constants $ \qql{1} $ and $ \qql{2} $ (further comments on the three space dimensional formulation will be given in Sec. \ref{sec-GK}). To reflect this, in the one space dimensional version, we also stick to this structure in \re{Ons-x_1-j(q,C)-1}--\re{Ons-x_1-j(q,C)-2}.

Now let us insert \re{Ons-x_1-j(q,C)-2} into \re{Ons-x_1-j(q,C)-1}. Then we obtain a generalized---non-Fourier type---constitutive relationship for heat current density---the Ginzburg--Landau-type relationship---, which reads
\begin{align}
    \label{q-CH}
    \qqq &= \qql{1} \pdr \frac{1}{\qT} + \qql{1} \qql{2} \pdr^2 \qqq .
\end{align}
Substituting equations \re{Ons-x_1-j(q,C)-1} and \re{Ons-x_1-j(q,C)-2} into \re{s-pr-x_1-j(q,C)}, we obtain
\begin{align}
    \qSig = \qql{1} \left[ \pdr  \left( \frac{1}{T} + \qql{2} \pdr \qqq \right) \right]^2 + \qql{2} \left( \pdr \qqq \right)^2 .
\end{align}
From this entropy production rate density, it seems rather nontrivial to construct a dissipation potential. In other words, assuming a one space dimensional state space, this model seems very difficult to be embedded in the GENERIC formulation. More closely, rewriting equation \re{q-CH} in the form
\begin{align}
    \left(1 - \qql{1} \qql{2} \pdr^2 \right) \qqq &= \qql{1} \pdr \frac{1}{\qT} 
\end{align}
and acting on \re{bal:e} by the operator $ \left(1 - \qql{1} \qql{2} \pdr^2 \right) $, we obtain
\begin{align}
    \qrho \pdt \qe - \qrho \qql{1} \qql{2} \pdt \pdr^2 \qe = - \qql{1} \pdr^2 \frac{1}{\qT} ,
\end{align}
which is difficult to formulate within GENERIC because dissipative coefficients appear on both sides of the equation. 

Indeed, in GENERIC one can have a dissipation potential on the right hand side, which is the usual form, or on the left hand side, acting on the time-derivative of a state variable, 
\begin{equation}
    \frac{\delta \Xi^*}{\delta \partial_t e} = e^* = \frac{1}{T}.
\end{equation}
Such a dual gradient dynamics is generated by a dual dissipation potential is obtained by Legendre transformation of the usual dissipation potential \cite{Janecka2018} and can be advantageous for instance when the dissipation potential is not smooth (e.g. in plasticity).

In the Ginzburg-Landau heat conduction, however, dissipative coefficients are both both sides. Therefore, in order to formulate that model within GENERIC, we would have to split to dual dissipation potential as $\Xi^* = \Xi_1^* \circ \Xi^*_2$ such that 
\begin{equation}
    \frac{\delta \Xi_2^*}{\delta \partial_t e} = \frac{\delta \Xi_1}{\delta e^*}
\end{equation}
with 
\begin{equation}
    \Xi_1(e^*) = \int_{L} \frac{l_1}{2} (\pdr e^*)^2 \dd r
    \qquad\text{and}\qquad
    \Xi^*_2(\dot{e}) = \int_{L} \frac{\qrho}{2} \left[\dot{e}^2 + \qql{2} \qql{1} (\pdr \dot{e})^2 \right] \dd r.
\end{equation}
However, to find the adjoint potential $\Xi_2^*$, one would have to solve a partial differential equation, dependent also on the boundary conditions.
On the other hand, we will show that with model reduction the Ginzburg--Landau type heat conduction can also be generated in the GENERIC framework.

\subsection{Two-dimensional state space}

Stepping forward, now let us assume that, besides our usual state variable $ e $, a further specific extensive vectorial state variable---denoted by $ \qqy $---is required to span the state space, \ie
\begin{align}
    \label{x_2^S}
    \qqx_2^S := \begin{pmatrix} \qe \\ \qqy \end{pmatrix} ,
\end{align}
for which the energy and entropic representations of the Gibbs relation are
\begin{subequations}
\begin{align}
    \label{Gibbs:e-x_2}
    \dd \qe &= \qT \dd \qs + \qqz \cdot \dd \qqy , \\
    \label{Gibbs:s-x_2}
    \dd \qs &= \frac{1}{\qT} \dd \qe - \frac{\qqz}{\qT} \cdot \dd \qqy ,
\end{align}
\end{subequations}
where the energy conjugated intensive function $ \qqz $ and the entropy conjugated intensive function $ - \frac{\qqz}{\qT} $ are defined by the partial derivatives $ \pder{\qe}{\qqy}{\qs} $ and $ \pder{\qs}{\qqy}{\qe} $, respectively. We can think of $ \qqy $ as an internal variable in the NET-IV sense, with the difference that, usually, NET-IV does not introduce and utilize the corresponding conjugate variable but works solely in terms of the internal variable,
shifting entropy by a concave---in the simplest cases a quadratic---function of the internal variable.
For instance formula (8) of \cite{Van2012} gives the relationship
\begin{align}
    \label{ext-s}
    \qs \left( \qe , \qqy \right) = \qs_0 \left( \qe \right) - \frac{ \qtilde m}{2} \qqy \cdot \qqy
\end{align}
with $ \pder{\qs}{\qe}{\qqy} = \frac{\dd \qs_0}{\dd \qe} = \frac{1}{\qT}$ and a non-negative $ \qtilde m $---a constant for simplicity. By deducing the differential of \re{ext-s} and comparing it with \re{Gibbs:s-x_2}, one can read off the constitutive relationship between $ \qqy $ and $ \qqz $ in this simplest case:
\begin{align}
    \label{z(y)}
    \frac{\qqz}{T} = \qtilde  m \qqy .
\end{align}
Later we will show that without such a constitutive relationship one cannot eliminate the internal variable(s) from the resulting system of equations.

\subsubsection{Jeffrey's type heat conduction}

Assuming the two-component state vector \re{x_2^S} and the usual entropy current density $ \qqJJ = \frac{1}{\qT} \qqq $, we obtain for entropy production rate density
\begin{align}
    \qSig = \qrho \pdt \qs + \pdr \qqJJ \stackrel{\re{Gibbs:s-x_2}}{=} \frac{1}{\qT} \qrho \pdt \qe - \frac{\qqz}{\qT} \qrho \pdt \qqy + \frac{1}{\qT} \pdr \qqq + \qqq \pdr \frac{1}{\qT} \stackrel{\re{bal:e}}{=} - \frac{\qqz}{\qT} \qrho \pdt \qqy + \qqq \pdr \frac{1}{\qT} \ge 0 ,
\end{align}
the positive semi-definiteness of which is ensured by the linear Onsagerian equations
\begin{subequations}
\begin{align}
    \label{Ons-x_2-1}
    \qqq &= \qql{11} \pdr \frac{1}{\qT} + \qql{12} \left( - \frac{\qqz}{\qT} \right) , \\
    \label{Ons-x_2-2}
    \qrho \pdt \qqy &= \qql{21} \pdr \frac{1}{\qT} + \qql{22} \left( - \frac{\qqz}{\qT} \right) ,
\end{align}
\end{subequations}
with the requirements
\begin{align}
    &&
    \qql{11} &\ge 0 , &
    \qql{22} & \ge 0, &
    \qql{11}\qql{22} - \qql{12}\qql{21} &\ge 0
    &&
\end{align}
on the coefficients $ \qql{ij} $. Equations \re{Ons-x_2-1} and \re{Ons-x_2-2} can be rewritten in the forms
\begin{align}
    \label{Ons-x_2-1-a}
    \qqq &= \qla \left( - \frac{\qqz}{\qT} \right) + \qql{11} \pdr \frac{1}{\qT} + \qls \left( - \frac{\qqz}{\qT} \right) , \\
    \label{Ons-x_2-2-a}
    \qrho \pdt \qqy &= -\qla \pdr \frac{1}{\qT} + \qls \pdr \frac{1}{\qT} + \qql{22} \left( - \frac{\qqz}{\qT} \right) , 
\end{align}
where the terms with the coefficients $ \qla := \left( \qql{12} - \qql{21} \right) / 2 $ do not increase the entropy while only the terms with the coefficients $ \qql{11} $, $ \qql{22} $ and $ \qls := \left( \qql{12} + \qql{21} \right) / 2 $ do. This can be seen from the conveniently rewritten entropy production rate density:
\begin{align}
    \label{s-pr-x_2}
    \qSig = \qql{11} \left( \pdr \frac{1}{\qT} \right)^2 + 2 \qls \left( \pdr \frac{1}{\qT} \right) \left( - \frac{\qqz}{\qT} \right) + \qql{22} \left( - \frac{\qqz}{\qT} \right)^2 .
\end{align}

In order to eliminate the internal variable $\qqy$ and its conjugate $\qqz$, we have to assume a constitutive relationship among these quantities. Let this relation be \re{z(y)}. Substituting it into \re{Ons-x_2-1} and \re{Ons-x_2-2}, we obtain
\begin{subequations}
\begin{align}
    \label{Ons-x_2-1-b}
    \qqq &= \qql{11} \pdr \frac{1}{\qT} + \qql{12} \left( - \qtilde m \qqy \right) , \\
    \label{Ons-x_2-2-b}
    \qrho \pdt \qqy &= \qql{21} \pdr \frac{1}{\qT} + \qql{22} \left( - \qtilde m \qqy \right) ,
\end{align}
\end{subequations}
the latter being rewritable as
\begin{align}  \label{ehehe}
    \left( \qrho \pdt + \qql{22} \qtilde  m \right) \qqy = \qql{21} \pdr \frac{1}{\qT} .
\end{align}
Letting the operator $ \left( \qrho \pdt + \qql{22} \qtilde  m \right) $ [observed on the LHS of \re{ehehe}] act on \re{Ons-x_2-1-b}, we find a thermodynamic version of the Jeffreys-type equation (originally proposed by Sir Harold Jeffreys in \cite{Jeffreys1929} regarding viscoelastic fluids -- for heat conduction, the model has been introduced by Joseph and Preziosi \cite{Joseph1989,Joseph1990}),
\begin{align}
    \tau \pdt \qqq + \qqq = \qlam_1 \pdr \frac{1}{\qT} + \qlam_2 \pdt \pdr \frac{1}{\qT} 
\end{align}
with the coefficients $ \tau := \frac{\qrho}{\qql{22} \qtilde  m } $, $ \qlam_1 := \frac{\det \qql{ij}}{ \qql{22}} $ and $ \qlam_2 := \frac{\qrho \qql{11}}{\qql{22} \qtilde  m} = \qql{11} \tau $.

Let us now give the GENERIC representation of this model. To obtain the reversible contribution of the time evolution, we use the energetic representation with the variables
\begin{align}
    \qqxi_2^\qE = \begin{pmatrix} \qsig \\ \qqeta \end{pmatrix} ,
\end{align}
where $\qqeta$ is the density of the internal variable $\qqy$, \ie $\qqeta = \qrho \qqy$. Since we neglect all mechanical interactions, the total energy equals to the internal energy, \ie
\begin{align}
    \qE = \int_V \qeps \left( \qqxi_2^\qE \right) \dd V.
\end{align}
Supposing that there are no non-local terms in the energy, its functional derivative \wrt to the state vector $\qqxi_2^\qE$ is
\begin{align}
    \var{\qE}{\qqxi_2^\qE} = \begin{pmatrix} T \\ \qqz \end{pmatrix}
\end{align}
[\cf the energetic representation of the Gibbs-relation \re{Gibbs:e-x_2}]. Assuming a Hamiltonian structure, the time evolution of the state variables under the reversible dynamics can be given as
\begin{align}
    \begin{pmatrix} \pdt \qsig \\ \pdt \qqeta \end{pmatrix} = 
    \begin{pmatrix} 0 & -\pdr \\ -\pdr & 0 \end{pmatrix} \begin{pmatrix} T \\ \qqz \end{pmatrix} ,
\end{align}
where the Poisson bivector is an antisymmetric operator (one can for instance construct the Poisson bracket to verify that) and it is constant, so it automatically satisfies the Jacobi identity.
The reversible evolution of the internal energy density is then
\begin{align}
    \pdt \qeps |_{\rm rev} = \pder{\qeps}{\qsig}{\qqeta} \pdt \qsig |_{\rm rev} + \pder{\qeps}{\qqeta}{\qsig} \pdt \qqeta |_{\rm rev} = - \qT \pdr \qqz - \qqz \pdr \qT = - \pdr \left( \qT \qqz \right) ,
\end{align}
and the energy is clearly conserved, which is a consequence of the above mentioned antisymmetry.

Such a Hamiltonian mechanics can be obtained by reduction of the Hamilton canonical equations for the cotangent bundle for fields \cite{Peshkov2018}. Therefore, since entropy is even (not affected) with respect to the time-reversal transformation (TRT), the other field $\qqz$ is odd. The Onsager-Casimir reciprocal relations then tell that the coupling between those variables via an operator antisymmetric with respect to the simultaneous adjugation and time-reversal, see e.g. \cite{Pavelka2018}. Since that operator is constant (consisting only of gradients) here, it has to be antisymmetric, which is in agreement with the structure of the operator here employed. Hamiltonian mechanics is compatible with the Onsager-Casimir reciprocal relations, and $\qqz$ is an odd quantity with respect to TRT.

To formulate the irreversible contribution to the time evolution, we switch to the entropic representation with the state vector
\begin{align}
    \qqxi_2^\qS = \begin{pmatrix} \qeps \\ \qqeta \end{pmatrix} .
\end{align}
Then entropy is
\begin{align}
    \qS = \int_\qV \qsig \left( \qqxi_2^\qS \right) \dd \qV
\end{align}
and the entropy conjugate state variables are
\begin{align}
    \qqxi_2^\qS{}^{^{^*}} = \var{\qS}{\qqxi_2^\qS} = \begin{pmatrix} \frac{1}{\qT} \\ - \frac{\qqz}{\qT} \end{pmatrix} .
\end{align}

Entropy production rate density \re{s-pr-x_2} is given in just these variables, thus the dissipation potential
\begin{align}
    \qXi_{\rm J} \left( \qqxi_2^\qS{}^{^{^*}} \right) = \int_\qL \left\{ \frac{\qql{11}}{2} \left( \pdr \frac{1}{\qT} \right)^2 + \frac{\qls}{2} \left[ \left( \pdr \frac{1}{\qT} \right) \left( - \frac{\qqz}{\qT} \right) - \frac{1}{\qT} \pdr \left( - \frac{\qqz}{\qT} \right) \right] + \frac{\qql{22}}{2} \left( - \frac{\qqz}{\qT} \right)^2 \right\} \dd \qr
\end{align}
can be introduced. The evolution of $\qqxi_2^\qS{}^{^{^*}}$ restricted to the irreversible dynamics is then
\begin{align}\label{eq.vec.irr}
    \left. \pdt \qqxi_2^\qS{}^{^{^*}} \right|_{\rm irr} = \begin{pmatrix} \pdt \qeps |_{\rm irr} \\[1.5ex] \pdt \qqeta |_{\rm irr} \end{pmatrix} = \left. \var{\qXi_{\rm J}}{\qqxi_2^\qS{}^{^{^*}}} \right|_{\qqxi_2^\qS{}^{^{^*}} = \var{\qS}{\qqxi_2^\qS}} = \begin{pmatrix} - \qql{11} \pdr^2 \frac{1}{\qT} - \qls \pdr \left( - \frac{\qqz}{\qT} \right) \\[1.5ex] \qls \pdr \frac{1}{\qT} + \qql{22} \left( - \frac{\qqz}{\qT} \right) \end{pmatrix} .
\end{align}
Here we can see coupling between $\frac{1}{T}$, which is even with respect to TRT, and $-\qqz/T$, which is odd. The Onsager-Casimir reciprocal relations then tell that the coupling is provided by an operator antisymmetric \wrt the simultaneous transposition (adjugation) and TRT. Therefore, the coefficients $\qls$ must be odd with respect to TRT. In particular they can not be constant, but the must depend on an odd quantity, here $\qqeta$ or $\qqz$. In other words, the dissipation potential has to be even \wrt TRT, which requires that the coefficients $\qls$ be odd.

Calculating the irreversible contribution to time evolution of entropy density yields
\begin{align}
    \nonumber
    \pdt \qsig |_{\rm irr} &= \pder{\qsig}{\qeps}{\qqeta} \pdt \qeps |_{\rm irr} + \pder{\qsig}{\qqeta}{\qeps} \pdt \qqeta |_{\rm irr} = -\pdr \left\{ \frac{1}{\qT} \left[ \qql{11} \pdr \frac{1}{\qT} + \qls \left( - \frac{\qqz}{\qT} \right) \right] \right\} \\
    \label{macska}
    & \hskip 10ex + \qql{11} \left( \pdr \frac{1}{\qT} \right)^2 + 2\qls \left( \pdr \frac{1}{\qT} \right) \left( - \frac{\qqz}{\qT} \right) + \qql{22} \left( - \frac{\qqz}{\qT} \right)^2 .
\end{align}

The final time evolution equations are the sum of the reversible and irreversible contributions, 
\begin{subequations}
\begin{align}
    \label{Jeff-dedt}
    \pdt \qeps &= - \pdr \left[ \qT \qqz + \qql{11} \pdr \frac{1}{\qT} + \qls \left( - \frac{\qqz}{\qT} \right) \right] , \\
    \label{Jeff-dsdt}
    \pdt \qsig &= - \pdr \left\{ \qqz + \frac{1}{\qT} \left[ \qql{11} \pdr \frac{1}{\qT} + \qls \left( - \frac{\qqz}{\qT} \right) \right] \right\} + \qql{11} \left( \pdr \frac{1}{\qT} \right)^2 + 2\qls \left( \pdr \frac{1}{\qT} \right) \left( - \frac{\qqz}{\qT} \right) + \qql{22} \left( - \frac{\qqz}{\qT} \right)^2 , \\
    \label{Jeff-dydt}
    \pdt \qqeta &= -\pdr \qT + \qls \pdr \frac{1}{\qT} + \qql{22} \left( - \frac{\qqz}{\qT} \right) .
\end{align}
    Note, however, that only two of them are independent while the equation for entropy or energy can be seen as an implication of the remaining two equations. One can also see that the part of the right hand side of these equations that transforms under TRT as the left hand side is coming from the Hamiltonian (reversible) dynamics while the part that flips sign under TRT \wrt the left hand side is coming form the dissipation potentials. 
\end{subequations}

Comparing \re{Jeff-dedt} with \re{bal:e} and \re{Jeff-dsdt} with \re{bal:s}, the heat and entropy current densities
\begin{align}
    \label{Jeff:q,J}
    \qqq &= \qT \qqz + \qql{11} \pdr \frac{1}{\qT} + \qls \left( - \frac{\qqz}{\qT} \right) , &
    \qqJJ &= \qqz + \frac{1}{\qT} \left[ \qql{11} \pdr \frac{1}{\qT} + \qls \left( - \frac{\qqz}{\qT} \right) \right]
\end{align}
can be recognized and the relationship
\begin{align}
    \qqJJ = \frac{1}{\qT} \qqq
\end{align}
is valid between them. The first expression of \re{Jeff:q,J} is equal to \re{Ons-x_2-1-a} in the approximation $ \frac{\qla}{\qT^2} \approx -1 $, then \re{Jeff-dydt}, and then \re{Ons-x_2-1-b} are also equivalent, thus full match is found between the equations generated by the two methodologies.

\subsubsection{Maxwell--Cattaneo--Vernotte heat conduction} \label{sec-MCV}

The Maxwell--Cattaneo--Vernotte equation is usually derived in the way that, in addition to specific internal energy, heat current density is also assumed as a state variable, and specific entropy is usually assumed in the form
\begin{align}
    \label{s(e,q)}
    \qhat \qs \left( \qe , \qqq \right) = \qs \left( \qe \right) - \frac{m}{2 \qrho} \qqq \cdot \qqq
\end{align}
with $ \pder{\qhat \qs}{\qe}{\qqq} = \frac{1}{\qT} $ and the positive constant $ m $.
(Here, the presence of $\qrho$ is to show that we respect that $\qqq$ is a specific extensive quantity.)
Constructing the differential of this expression, we obtain
\begin{subequations}
\begin{align}
    \dd \qhat \qs = \frac{1}{\qT} \dd \qe - \frac{m}{\qrho} \qqq \cdot \dd \qqq , 
\end{align}
and its rearranged form
\begin{align}
    \dd \qe = \qT \dd \qhat \qs + \frac{m}{\qrho} \qT \qqq \cdot \dd \qqq ,
\end{align}
\end{subequations}
which tells us that specific internal energy depends on (extended) entropy and heat current density both. When thinking of $\qe$ and $\qqq$ as the temporal and spatial parts of a single internal energy spacetime
quantity, it appears highly inconsistent that the temporal part of a spacetime quantity depends on the spatial part of the same spacetime quantity.

Consequently, instead of this assumption, let us get back to equations \re{Ons-x_2-1-a} and let us identify heat current density with the entropy conjugated internal variable as
\begin{align}
    \label{id:q=z}
    \qqq = \qla \left( - \frac{\qqz}{\qT} \right) ,
\end{align}
which means that $ \qql{11} = 0 $ and $ \qls = 0 $. Then \re{Ons-x_2-2-a} reduces to
\begin{align}
    \qrho \pdt \qqy &= -\qla \pdr \frac{1}{\qT} + \frac{\qql{22}}{\qla} \qqq , 
\end{align}
and, substituting the constitutive equation \re{z(y)}, we obtain
\begin{align}
    - \frac{\qrho}{\qtilde m \qla} \pdt \qqq &= -\qla \pdr \frac{1}{\qT} + \frac{\qql{22}}{\qla} \qqq .
\end{align}
Introducing the coefficients $ \ql := \qql{22} = \qla $ and $ \tau := \frac{\qrho}{\qtilde m \ql} $, one can recognise the well-known Maxwell--Cattaneo--Vernotte equation
\begin{align}
    \label{MCV}
    \tau \pdt \qqq + \qqq = \ql \pdr \frac{1}{\qT} .
\end{align}
Let us make here two remarks.
    Firstly, to obtain the MCV equation, the usual entropy current density relationship $ \qqJJ = \frac{1}{\qT} \qqq $ has been assumed.
    Secondly, the heat current density has been identified with the entropy conjugate internal variable (apart from a constant multiplier), but not with the internal variable itself. When a simple linear relationship \re{z(y)} is assumed between the internal variable and its conjugate then the identification between the heat current density and the internal variable also holds. More generally, however, the situation is more complicated. When deriving heat conduction models in GENERIC, the step of identification is a key ingredient, see \eg in \cite{Lebon2017} and \cite{Grmela2019}.

For the GENERIC treatment of MCV model, let us refer on \cite{Grmela2019}. Due to microscopic considerations, the vectorial variable is connected to the momentum of phonons and, correspondingly, the density of the internal variable $\qqeta$ is identified with the conjugate of entropy current density\footnote{As far as the reversible part is concerned $ \qqw $ is typically conjugate to the entropy flux. However, if the dissipation potential contains some nonlocal terms, then entropy current density gets additional terms, we will show this \eg in Sec. \re{sec-GK}. So calling $ \qqw $ as conjugate entropy flux is completely all right when we talk about the reversible evolution, but more precisely, conjugate $ \qqw $ is a part of the entropy flux. Furthermore, $ \qqw $ can be identified with the momentum density of phonons divided by the entropy density, so it could also be called specific phonon momentum \cite{Pavelka2018}.} $\qqw$. The reversible time evolution of the variables
\begin{align}
    \qhat \qqxi_2^\qE = \begin{pmatrix} \qsig \\ \qqw \end{pmatrix} 
\end{align}
is given as
\begin{align}
    \begin{pmatrix} \pdt \qsig |_{\rm rev} \\ \pdt \qqw |_{\rm rev} \end{pmatrix} = 
    \begin{pmatrix} 0 & -\pdr \\ -\pdr & 0 \end{pmatrix} \begin{pmatrix} T \\ \qqJ \end{pmatrix} .
\end{align}
The irreversible contribution to time evolution is derived in the variables of entropic representation
\begin{align}
    \qhat \qqxi_2^\qS = \begin{pmatrix} \qeps \\ \qqw \end{pmatrix} .
\end{align}
Dissipation potential is assumed in the entropy conjugated variables
\begin{align}
    \qhat \qqxi_2^\qS{}^{^{^*}} = \var{\qS}{\qqxi_2^\qS} = \begin{pmatrix} \frac{1}{\qT} \\ - \frac{\qqJ}{\qT} \end{pmatrix}
\end{align}
with the form
\begin{align}
    \label{dis-pot-MCV}
    \qXi_{\rm MCV} \left( \qhat \qqxi_2^\qS{}^{^{^*}} \right) = \int_\qL \frac{\qk}{2} \left( - \frac{\qqJ}{\qT} \right)^2 \dd \qr ,
\end{align}
where $\qk$ is a non-negative constant. The generated gradient dynamics of $ \qhat \qqxi_2^\qS{}^{^{^*}}$ is then
\begin{align}
    \left. \pdt \qhat \qqxi_2^\qS{} \right|_{\rm irr} = \begin{pmatrix} \pdt \qeps |_{\rm irr} \\[1ex] \pdt \qqw |_{\rm irr} \end{pmatrix} = \left. \var{\qXi_{\rm MCV}}{\qhat \qqxi_2^\qS{}^{^{^*}}} \right|_{\qhat \qqxi_2^\qS{}^{^{^*}} = \var{\qS}{\qhat \qqxi_2^\qS}} = \begin{pmatrix} 0 \\ \qk \left( - \frac{\qqJ}{\qT} \right) \end{pmatrix} ,
\end{align}
from which the irreversible contribution to the time evolution of entropy density is
\begin{align}
    \pdt \qsig |_{\rm irr} = \qk \left( - \frac{\qqJ}{\qT} \right)^2 .
\end{align}
Finally, the system of time evolution equations is
\begin{subequations}
\begin{align}
    \label{MCV-GENERIC-s}
    \pdt \qsig &= - \pdr \qqJ + \qk \left( - \frac{\qqJ}{\qT} \right)^2 , \\
    \label{MCV-GENERIC}
    \pdt \qqw &= - \pdr \qT + \qk \left( - \frac{\qqJ}{\qT} \right) ,
\end{align}
\end{subequations}
comparing \re{MCV-GENERIC-s} with \re{bal:s} the entropy current density $ \qqJJ = \qqJ $ can be identified. Time evolution of internal energy density is then
\begin{align}
    \pdt \qeps = \pdt \qeps |_{\rm rev} + \pdt \qeps |_{\rm irr} =  - \pdr \left( \qT \qqJ \right) + 0 = - \pdr \left( \qT \qqJ \right) ,
\end{align}
and, comparing this with \re{bal:e}, the heat current density -- entropy current density relationship
\begin{align}  \label{qTjjTq}
    && && &&
    \qqq &= \qT \qqJJ &&
    \Longleftrightarrow &
    \qqJJ &= \frac{1}{\qT} \qqq
    && && &&
\end{align}
can be read off. Assuming a linear relationship between $\qqJ$ and $\qqw$, in the linear approximation the equivalence of \re{MCV-GENERIC} can \re{MCV} be recognized.

\subsubsection{Model reduction: Fourier heat conduction}

Now let us assume that the equilibration time of $\qqw$ is much shorter than that of $\qs$. In this approximation, \re{MCV-GENERIC} reduces to
\begin{align}
    0 = - \pdr \qT + \qk \left( - \frac{\qqJ}{\qT} \right) ,
\end{align}
which can be treated as a constitutive equation among $\qqJ$ and $\qT$, moreover, it is (also in the linear approximation) equal to Fourier's law.

\subsubsection{Guyer--Krumhansl heat conduction} \label{sec-GK}

In the framework of NET-IV, the GK equation is derived with the assumptions of an extended specific entropy in the form of \re{s(e,q)} and of a generalized entropy current density as defined in \re{j(q,C)}. With these, the entropy production rate density is
\begin{align}
    \label{s-pr-GK}
    \qSig = \qqq \qodot \left( - m \pdt \qqq + \pdr \frac{1}{\qT} + \pdr \qqC \right) + \qqC \qodot \pdr \qqq .
\end{align}
The linear equations
\begin{align}
    \label{Ons-x_2-1-GK}
    \qqq &= \qql{1} \left( - m \pdt \qqq + \pdr \frac{1}{\qT} + \pdr \qqC \right) , \\
    \label{Ons-x_2-2-GK}
    \qqC &= \qql{2} \pdr \qqq
\end{align}
with non-negative constants $\qql{1}$ and $\qql{2}$ ensure the positive semi-definiteness of the entropy production rate density \re{s-pr-GK}. Inserting \re{Ons-x_2-2-GK} into \re{Ons-x_2-1-GK}, we obtain the Guyer--Krumhansl equation
\begin{align}
    \label{GK}
    \tau \pdt \qqq + \qqq = \qql{1} \pdr \frac{1}{\qT} + \kappa^2 \pdr^2 \qqq 
\end{align}
with the coefficients $ \tau := \qql{1} m $ and $ \kappa^2 := \qql{1} \qql{2} $.

As for GENERIC, since the GK equation can be considered as a nonlocal extension of the MCV model, the reversible dynamics needs no modification, but it is only the dissipation potential defined in \re{dis-pot-MCV} that has to be extended by a non-local term, \viz
\begin{align}
    \qXi_{\rm GK} \left( \qhat \qqxi_2^\qS{}^{^{^*}} \right) = \qXi_{\rm MCV} \left( \qhat \qqxi_2^\qS{}^{^{^*}} \right) + \int_\qL \frac{\qk_2}{2} \left[ \pdr  \left( - \frac{\qqJ}{\qT} \right) \right]^2 \dd \qr = \int_\qL \left\{ \frac{\qk_1}{2} \left( - \frac{\qqJ}{\qT} \right)^2 + \frac{\qk_2}{2} \left[ \pdr  \left( - \frac{\qqJ}{\qT} \right) \right]^2 \right\} \dd \qr ,
\end{align}
where $\qk_1$ and $\qk_2$ are non-negative constants. Consequently, the GENERIC representation of GK heat conduction is
\begin{align}
    \pdt \qeps &= - \pdr \left( \qT \qqJ \right) , \\
    \label{kutya}
    \pdt \qsig &= - \pdr \left[ \qqJ + \qk_2 \left( - \frac{\qqJ}{\qT} \right) \pdr \left( - \frac{\qqJ}{\qT} \right) \right] + \qk_1 \left( - \frac{\qqJ}{\qT} \right)^2 + \qk_2 \left[ \pdr  \left( - \frac{\qqJ}{\qT} \right) \right]^2 , \\
    \label{GK-GENERIC}
    \pdt \qqw &= - \pdr \qT + \qk_1 \left( - \frac{\qqJ}{\qT} \right) - \qk_2 \pdr^2 \left( - \frac{\qqJ}{\qT} \right) .
\end{align}
Now we can recognise heat and entropy current densities
\begin{align}
    \qqq &= \qT \qqJ & \text{and} &&
    \qqJJ &= \qqJ + \qk_2 \left( - \frac{\qqJ}{\qT} \right) \pdr \left( - \frac{\qqJ}{\qT} \right) .
\end{align}
In an appropriate approximation, and utilizing what we had in and below \re{qTjjTq} for $\qqw$, $\qqJ$, and $\qqq$, \re{GK-GENERIC} proves equivalent to \re{GK} (see more on it in \cite{Grmela2019}) and, in this approximation the entropy current density -- heat current density relationship
\begin{align}
    \label{GK:q-J}
    \qqJJ &= \frac{1}{T} \qqq + \qk_2 \left( - \frac{\qqq}{\qT^2} \right) \pdr \left( - \frac{\qqq}{\qT^2} \right) \approx \left( \frac{1}{\qT} + \tilde \qk_2 \pdr \qqq \right) \cdot \qqq
\end{align}
can be read, which is equivalent with \re{j(q,C)} by substituting \re{Ons-x_2-2-GK}.

\subsubsection{Comments on the 3D GK equation}

At this point, let us make a detour by outlining the whole three-dimensional counterpart of the one dimensional treatment, especially because the 3D GK equation finds particularly many applications, especially in isotropic materials.

In the NET-IV approach, entropy production rate density is
\begin{align}
    \qSig = \qqq \cdot \left( - m \pdt \qqq + \nablar \frac{1}{\qT} + \qqC \cdot \nablal \right) + \qqC \qodot \left( \qqq \nablal \right) ,
\end{align}
and the Onsagerian equations are
\begin{align}
    \qqq &= \ql \left( - m \pdt \qqq + \nablar \frac{1}{\qT} + \qqC \cdot \nablal \right) , \\
    \qqC &= \bbL \qodot \left( \qqq \nablal \right) ,
\end{align}
where $\ql$ is a non-negative constant and $\bbL$ is a constitutive fourth-order isotropic tensor,
\begin{align}
    \label{repr4th}
    (\bbL)_{ijkl} = \frac{\sfL^{\rm sph}-\sfL^{\rm dev}}{3} \delta_{ij} \delta_{kl} + \frac{\sfL^{\rm dev}+\sfL^{\rm A}}{2} \delta_{ik} \delta_{jl} + \frac{\sfL^{\rm dev}-\sfL^{\rm A}}{2} \delta_{il} \delta_{jk} ,
\end{align}
where $ \delta $ denotes the Kronecker delta and $ \sfL^{\rm dev} \ge 0 $ , $ \sfL^{\rm sph} \ge 0 $ and $ \sfL^{\rm A} \ge 0 $ are scalar coefficients (related to the symmetric deviatoric, spherical, and antisymmetric deviatoric parts of second-order tensors, respectively). As a result, the (three spatial dimensional form of the) GK equation follows:
\begin{align}
    \tau \pdt \qqq + \qqq = - \ql \nablar \frac{1}{\qT} + \mu_1
    \triangler
    \qqq + \mu_2 \nablar \left( \nablar \cdot \qqq \right)
\end{align}
with the coefficients $ \tau = \ql \qtilde m $, $ \mu_1 = \ql \frac{\sfL^{\rm dev}+\sfL^{\rm A}}{2} $ and $ \mu_2 = \ql \frac{2 \sfL^{\rm sph} + \sfL^{\rm dev} - 3 \sfL^{\rm A}}{6}$.

In the GENERIC formulation, correspondingly to the increasing tensorial order of the state variables $ \qhat \qqxi_2^\qE =
\begin{pmatrix} \qsig \\ \qqw \end{pmatrix}
$, the antisymmetric operator matrix will change to
\begin{align}
    \qqL =
    \begin{pmatrix}
    0 & - \nablar \cdot \\
    - \nablar  & 0 
    \end{pmatrix} 
\end{align}
and the dissipation potential is sought for in the form
\begin{align}
    \qXi_{\rm GK,3D} \left( \qhat \qqxi_2^\qS{}^{^{^*}} \right) = \int_\qL \left\{ \frac{\qk}{2} \left( - \frac{\qqJ}{\qT} \right) \cdot \left( - \frac{\qqJ}{\qT} \right) + \left( \frac{\bbK}{2} \qodot \left[ \left( - \frac{\qqJ}{\qT} \right)  \nablal \right]  \right) \qodot \left[ \left( - \frac{\qqJ}{\qT} \right) \nablal \right] \right\} \dd \qr ,
\end{align}
where $ \qk \ge 0 $ and $\bbK$ is a positive definite fourth-order tensor with an isotropic representation analogous to \re{repr4th}.

It is appropriate to comment here that, in \cite{Szucs2019}, the deviatoric and spherical parts of tensors are treated as independent state variables, making the discussion clearer in certain aspects. In subsection \ref{sec:3Dstsp}, a third-order tensors also appears. In general, third-order tensors are coupled via a sixth-order tensor, hence, tensorially complete calculations can become very complicated. For this reason, we do not give more remarks on the three spatial dimensional treatment but, for more details, we refer to \cite{Fama2021}.

\subsubsection{Model reduction: Ginzburg--Landau equation} \label{sec-red-CH}

When fast evolution of $\qqw$ is relaxed, equation \re{GK-GENERIC} reduces to
\begin{align}  \label{CHred}
    0 = - \pdr \qT + \qk_1 \left( - \frac{\qqJ}{\qT} \right) - \qk_2 \pdr^2 \left( - \frac{\qqJ}{\qT} \right) ,
\end{align}
in which the structure of the Ginzburg--Landau type equation \re{q-CH} can be recognised.

\subsubsection{Summary}
As a conclusion we find the equivalence of Jeffrey's, MCV and GK time evolution equations and the belonging entropy current density -- heat current density relationships. We have shown, that GL type heat conduction (which seems to be incompatible with GENERIC) can be obtained in the framework by reducing GK time evolution. We have seen that nonlocal terms in the dissipation potential give additional contributions to the entropy current density. In what follows, we extend our state space with an additional tensorial state variable, we derive BD heat conduction equation within NET-IV and GENERIC and we show that higher order tensorial terms in the entropy current density -- heat current density relationship or in the dissipation potential lead to such equations that can be seen as a reduced model of the dynamics in a higher dimensional state space.

\subsection{Three-dimensional state space} \label{sec:3Dstsp}

For describing heat conduction phenomena beyond the MCV and GK models, a further state variable is also required (which is a second-order tensor when treated in three space dimensions). In this case, the thermodynamical state space is spanned by the variables
\begin{align}
    \qqx_3^S := \begin{pmatrix} \qe \\ \qqq \\ \qqQ \end{pmatrix} ,
\end{align}
where $\qqQ$ is the additional internal variable and, furthermore, let us refer to Section \ref{sec-MCV}, where heat current density is understood as a state variable after the identification \re{id:q=z}, and a linear relationship among the internal variable and its conjugate holds.

\subsubsection{Ballistic--diffusive heat conduction} \label{sec-BC}

Let us assume that specific entropy \re{s(e,q)} is shifted by a concave expression of the new internal variable $\qqQ$ as well: 
\begin{align}
    \label{s(e,q,Q)}
    \qhat{\qhat \qs} \left( \qe , \qqq , \qqQ \right) = \qs \left( \qe \right) - \frac{m_1}{2 \qrho} \qqq \cdot \qqq - \frac{m_2}{2 \qrho} \qqQ \qodot \qqQ ,
\end{align}
where $m_1$ and $m_2$ are non-negative constants (for simplicity---in general, these can again be functions depending on the state variables) and that a generalized entropy current density relationship \re{j(q,C)} holds. Then entropy production rate density is
\begin{align}
    \label{s-pr-balcond}
    \qSig = \qqq \Cdot \left( -m_1 \pdt \qqq + \pdr \frac{1}{\qT} + \pdr \qqC \right) - m_2 \qqQ \qodot \pdt \qqQ + \qqC \qodot \pdr \qqq .
\end{align}
Here, the first term a vector--vector coupling while the second and third terms are tensor--tensor couplings (in a three space dimensional sense). Therefore, in the spirit of Curie's principle, the linear Onsagerian equations
\begin{subequations}
\begin{align}
    \label{Ons-x_3-1}
    - m_1 \pdt \qqq + \pdr \frac{1}{\qT} + \pdr \qqC &= \ql \qqq , \\
    \label{Ons-x_3-2}
    - m_2 \pdt \qqQ &= \qql{11} \qqQ + \qql{12} \pdr \qqq , \\
    \label{Ons-x_3-3}
    \qqC &= \qql{21} \qqQ + \qql{22} \pdr \qqq
\end{align}
\end{subequations}
are taken to ensure positive-semidefiniteness of \re{s-pr-balcond} with the coefficients fulfilling the requirements
\begin{align}
    \label{req:l}
    &&
    \ql &\ge 0 , &
    \qql{11} &\ge 0 , &
    \qql{22} & \ge 0, &
    \qql{11}\qql{22} - \qql{12}\qql{21} &\ge 0 .
    &&
\end{align}
Similarly to \re{Ons-x_2-1-a} and \re{Ons-x_2-2-a}, equations \re{Ons-x_3-2} and \re{Ons-x_3-3} can be rewritten into entropy-preserving and entropy-increasing contributions, \viz
\begin{align}
    \label{Ons-x_3-2-a}
    - m_2 \pdt \qqQ &=
    \left(
    - \qla \pdr \qqq
    \right)
    +
    \left( \qql{11} \qqQ + \qls \pdr \qqq
    \right)
    , \\    \label{Ons-x_3-3-a}
    \qqC &=
    \left( \qla \qqQ
    \right)
    +
    \left( \qls \qqQ + \qql{22} \pdr \qqq
    \right)
    .
\end{align}

After eliminating the entropy current density multiplier $\qqC$ from \re{Ons-x_3-1}, the time evolution equations
\begin{align}
    \label{d(q,Q)dt}
    \begin{pmatrix} - m_1 \pdt \qqq \\ - m_2 \pdt \qqQ \end{pmatrix} =
    \begin{pmatrix} - \pdr \frac{1}{\qT} - \qla \pdr \qqQ \\ - \qla \pdr \qqq \end{pmatrix} +
    \begin{pmatrix} \ql \qqq - \qls \pdr \qqQ - \qql{22} \pdr^2 \qqq \\ \qql{11} \qqQ + \qls \pdr \qqq \end{pmatrix}
\end{align}
follow. At this point, one has to recognize that, although the antisymmetric part has no contribution in the entropy production rate density, it influences the evolution equations. At first sight, this separation may look unusual and artificial but will be clarified later when presenting the derivation in the framework of GENERIC. The first term on the right hand side of \re{d(q,Q)dt} represents the reversible contribution, while the second term is the irreversible contribution of time evolution. Now writing the second equation of \re{d(q,Q)dt} in time-derivative operator form, namely,
\begin{align}
    \left( m_2 \pdt + \qql{11} \right) \qqQ = \left( \qla - \qls \right) \pdr \qqq ,
\end{align}
and letting the operator $\left( m_2 \pdt + \qql{11} \right)$ act on the first equation of \re{d(q,Q)dt}, the outcome is the extended constitutive equation of heat flux, \ie
\begin{align}
    m_1 m_2 \pdt^2 \qqq + \left( m_1 \qql{11} + m_2 \ql \right) \pdt \qqq - m_2 \qql{22} \pdt \pdr^2 \qqq -\left( \qql{11}\qql{22} - \qql{12}\qql{21} \right) \pdr^2 \qqq + \ql \qql{11} \qqq = m_2 \pdt \pdr \frac{1}{ \qT} + \qql{11} \pdr \frac{1}{\qT} .
\end{align}
We can observe that this equation is a non-local extension of the ballistic--diffusive heat conduction equation. If we assume that $\qql{22}=0$ then we obtain for \re{Ons-x_3-3}
\begin{align}
    \qqC = \qql{21} \qqQ ,
\end{align}
which means that the entropy current multiplier is identified with the internal variable, thus
\begin{align}
    \qqJJ = \left( \frac{1}{T} \ident + \qql{21} \qqQ \right) \qqq .
\end{align}
Then the first equation of \re{d(q,Q)dt} is
\begin{align}
    - m_1 \pdt \qqq = \left( - \pdr \frac{1}{\qT} - \qla \pdr \qqQ \right) + \left( \ql \qqq - \qls \pdr \qqQ \right)
\end{align}
and, finally, the constitutive equation of ballistic--diffusive heat conduction is
\begin{align}
    m_1 m_2 \pdt^2 \qqq + \left( m_1 \qql{11} + m_2 \ql \right) \pdt \qqq -\left( \qql{11}\qql{22} - \qql{12}\qql{21} \right) \pdr^2 \qqq + \ql \qql{11} \qqq = m_2 \pdt \pdr \frac{1}{ \qT} + \qql{11} \pdr \frac{1}{\qT} .
\end{align}

In the GENERIC framework, we look for the time evolution of the variables
\begin{align}
    \qqxi_3^\qE = \begin{pmatrix} \qsig \\ \qqw \\ \qqW \end{pmatrix}
\end{align}
(where $\qqW$ is a second-order tensorial variable---not
required to be symmetric and/or traceless%
---if treated in three space dimensions). The partial derivatives of internal energy density $\qeps \left( \qqxi_3^\qE \right)$ defines the energy conjugate variables
\begin{align}
    &&
    \qT &= \pder{\qeps}{\qsig}{\qqw,\qqW} , &
    \qqJ &= \pder{\qeps}{\qqw}{\qsig,\qqW} , &
    \qqP &= \pder{\qeps}{\qqW}{\qsig,\qqw} .
    &&
\end{align}
Time evolution of $\qqxi_3^\qE$ restricted to reversible dynamics is
\begin{align}
    \begin{pmatrix} \pdt \qsig |_{\rm rev} \\ \pdt \qqw |_{\rm rev} \\ \pdt \qqW |_{\rm rev} \end{pmatrix} = 
    \begin{pmatrix} 0 & -\pdr & 0 \\ -\pdr & 0 & - \pdr \\ 0 & -\pdr & 0 \end{pmatrix} \begin{pmatrix} T \\ \qqJ \\ \qqP \end{pmatrix} .
\end{align}
Why is the Poisson bivector chosen in this way here? In general, there are several routes towards the bivectors (or Poisson brackets): (i) analogy with quantum mechanical commutators, (ii) the theory of Lie groups and algebras, and (iii) reduction from a more detailed Poisson bracket. None of these routes has been taken in the present work, so the chosen Poisson bivector can only be seen as a suitable choice advantageous for the comparison with NET-IV. We believe, however, that it can be interpreted as an approximation of a more detailed and microscopically motivated Poisson bivector, as for instance the top left $ 2 \times 2 $ block can be obtained from the Poisson bivector of fluid mechanics when neglecting the momentum density. 

From the Onsager-Casimir reciprocal relations it again follows that the parity of $\qqW$ is even so that the coupling can be antisymmetric.

To formulate the irreversible contribution of time evolution, we switch to the variables of entropic representation: the dissipation potential is taken in the entropy conjugated variables
\begin{align}
    &&
    \qqxi_3^\qS &= \begin{pmatrix} \qeps \\ \qqw \\ \qqW \end{pmatrix} ,  & \Longrightarrow &&
    \qqxi_3^\qS{}^{^{^*}} &= \var{\qS}{\qqxi_3^\qS} = \begin{pmatrix} \frac{1}{\qT} \\ - \frac{\qqJ}{\qT_{}} \\ - \frac{\qqP^{}}{\qT} \end{pmatrix}
    &&
\end{align}
as the quadratic expression
\begin{align}
    \nonumber
    \qXi_{\rm BD} \left( \qqxi_3^\qS{}^{^{^*}} \right) &= \int_\qL \bigg\{ \frac{\qk}{2} \left( - \frac{\qqJ}{\qT} \right)^2 + \frac{\qk_{11}}{2} \left( - \frac{\qqP}{\qT} \right)^2 \\
    \label{dis-pot-BC}
    & \hskip 10ex + \frac{\qk_{12}^{\rm S}}{2} \left[ \left( - \frac{\qqP}{\qT} \right) \pdr \left( - \frac{\qqJ}{\qT} \right) - \left( - \frac{\qqJ}{\qT} \right) \pdr \left( - \frac{\qqP}{\qT} \right) \right] + \frac{\qk_{22}}{2} \left[ \pdr \left( - \frac{\qqJ}{\qT} \right) \right]^2 \bigg\} \dd \qr
\end{align}
with the convexity-ensuring requirements
\begin{align}
    &&
    \qk &\ge 0 , &
    \qk_{11} &\ge 0 , &
    \qk_{22} & \ge 0, &
    \qk_{11}\qk_{22} - \left( \qk_{12}^{\rm S} \right)^2 &\ge 0 .
    &&
\end{align}
The Onsager-Casimir reciprocal relation then require that $\qk$, $\qk_{11}$, and $\qk_{22}$ be even \wrt TRT while $\qk_{12}^{\rm S}$ be odd, e.g. a linear function of $\qqw$ or $\qqJ$.
This dissipation potential generates the irreversible contribution to the time evolution of $\qqxi_3^\qS$ as
\begin{align}
    \left. \pdt \qqxi_3^\qS{}^{^{^*}} \right|_{\rm irr} = \begin{pmatrix} \pdt \qeps |_{\rm irr} \\ \pdt \qqw |_{\rm irr} \\ \pdt \qqW |_{\rm irr} \end{pmatrix} = \left. \var{\qXi_{\rm BD}}{\qqxi_3^\qS{}^{^{^*}}} \right|_{\qqxi_3^\qS{}^{^{^*}} = \var{\qS}{\qqxi_3^\qS}} = \begin{pmatrix} 0 \\ \qk \left( - \frac{\qqJ}{\qT} \right) - \qk_{12}^{\rm S} \pdr \left( - \frac{\qqP}{\qT} \right) - \qk_{22} \pdr^2 \left( - \frac{\qqJ}{\qT} \right) \\ \qk_{11} \left( - \frac{\qqP}{\qT} \right) + \qk_{12}^{\rm S} \pdr \left( - \frac{\qqJ}{\qT} \right)  \end{pmatrix} .
\end{align}
The irreversible contribution to time evolution of entropy density is
\begin{align}
    \nonumber
    \pdt \qsig |_{\rm irr} &= - \pdr \left\{ \left[ \qk_{12}^{\rm S}  \left( - \frac{\qqP}{\qT} \right) + \qk_{22} \pdr \left( - \frac{\qqJ}{\qT} \right) \right] \left( - \frac{\qqJ}{\qT} \right) \right\} \\
    & \hskip 5ex + \qk \left( - \frac{\qqJ}{\qT} \right)^2 + \qk_{11} \left( - \frac{\qqP}{\qT} \right)^2 + 2 \qk_{12}^{\rm S} \left( - \frac{\qqP}{\qT} \right) \pdr \left( - \frac{\qqJ}{\qT} \right) + \qk_{22} \left[ \pdr \left( - \frac{\qqJ}{\qT} \right) \right]^2 ,
\end{align}
and, the time evolution of $\qqxi_3^\qE$ is
\begin{subequations}
\begin{align}
    \nonumber
    \pdt \qsig &= - \pdr \left\{ \qqJ + \left[ \qk_{12}^{\rm S}  \left( - \frac{\qqP}{\qT} \right) + \qk_{22} \pdr \left( - \frac{\qqJ}{\qT} \right) \right] \left( - \frac{\qqJ}{\qT} \right) \right\} \\
    & \hskip 5ex + \qk \left( - \frac{\qqJ}{\qT} \right)^2 + \qk_{11} \left( - \frac{\qqP}{\qT} \right)^2 + 2 \qk_{12}^{\rm S} \left( - \frac{\qqP}{\qT} \right) \pdr \left( - \frac{\qqJ}{\qT} \right) + \qk_{22} \left[ \pdr \left( - \frac{\qqJ}{\qT} \right) \right]^2 , \\
    \label{dwdt}
    \pdt \qqw &= - \pdr \qT - \pdr \qqP + \qk \left( - \frac{\qqJ}{\qT} \right) - \qk_{12}^{\rm S} \pdr \left( - \frac{\qqP}{\qT} \right) - \qk_{22} \pdr^2 \left( - \frac{\qqJ}{\qT} \right) , \\
    \label{dWdt}
    \pdt \qqW &= - \pdr \qqJ + \qk_{11} \left( - \frac{\qqP}{\qT} \right) + \qk_{12}^{\rm S} \pdr \left( - \frac{\qqJ}{\qT} \right) .
\end{align}
\end{subequations}

Comparing \re{dwdt} and \re{dWdt} with \re{d(q,Q)dt}, the same structures can be recognised. Naturally, if $\qk_{22}=0$, we simply omit the last term from the dissipation potential \re{dis-pot-BC} and obtain the ballistic--diffusive heat conduction. Time evolution of internal energy density is then
\begin{align}
    \pdt \qeps = - \qT \pdr \qqJ - \qqJ \left( \pdr \qT + \pdr \qqP \right) - \qqP \pdr \qqJ = - \pdr \left[ \left( \qT \ident + \qqP \right) \qqJ \right].
\end{align}
Heat current and entropy current densities 
\begin{align}
    \label{BD:q-J}
    \qqq &= \left( \qT \ident + \qqP \right) \qqJ , &
    \qqJJ &= \qqJ + \left[ \qk_{12}^{\rm S}  \left( - \frac{\qqP}{\qT} \right) + \qk_{22} \pdr \left( - \frac{\qqJ}{\qT} \right) \right] \left( - \frac{\qqJ}{\qT} \right)
\end{align}
can be read off. Expressing $ \qqJ $ with $ \qqq $ and replacing it to $ \qqJJ $ in an appropriate approximation entropy current density -- heat current density relationship obtained in \re{BD:q-J} is proved to be equivalent with \re{j(q,C)} by substituting \re{Ons-x_3-3}. This means that (apart from non-linear terms) we find a whole agreement between the equations derived on two ways. We remark that in the full 3D treatment the second order tensor $ \qT \ident + \qqP $ have to be inverted. Since $ \qT > 0 $, thus $ \qT \ident $ and $ \qT \ident + \qqP $ are nonsingular and, based on \cite{Miller1981} the inverse of $ \qT \ident + \qqP $ can be given as
\begin{align}
    \left( \qT \ident + \qqP \right)^{-1} = \frac{1}{\qT} \ident + \qqR 
\end{align}
with
\begin{align}
    \qqR = - \frac{1}{\qT^2} \left( \ident + \frac{1}{T} \qqP \right)^{-1} \qqP.
\end{align}

\subsubsection{Model reduction: Guyer--Krumhansl and Ginzburg--Landau equations}

Supposing $ \qk_{12}^{\rm S} = 0 $ and $\qk_{22} = 0$, \re{dis-pot-BC} is an algebraic dissipation potential and the time evolution equations are
\begin{subequations}
\begin{align}
    \pdt \qsig &= - \pdr \qqJ + \qk \left( - \frac{\qqJ}{\qT} \right)^2 + \qk_{11} \left( - \frac{\qqP}{\qT} \right)^2 , \\
    \label{dwdt-GK-GEN}
    \pdt \qqw &= - \pdr \qT - \pdr \qqP + \qk \left( - \frac{\qqJ}{\qT} \right) , \\
    \pdt \qqW &= - \pdr \qqJ + \qk_{11} \left( - \frac{\qqP}{\qT} \right) .
\end{align}
\end{subequations}
First of all, we can recognize that $ \qqJJ = \qqJ $ and from the balance of internal energy
\begin{align}
    \label{GK:q-JJ}
    \qqq = \left( \qT \ident + \qqP \right) \qqJ = \left( \qT \ident + \qqP \right) \qqJJ ,
\end{align}
Assuming that fast time evolution of $\qqW$ is over,
\begin{align}
    \label{rel:W}
    0 &= - \pdr \qqJ + \qk_{11} \left( - \frac{\qqP}{\qT} \right) ,
\end{align}
which can be treated as a constitutive relation among $\qqJ$ and $\frac{\qqP}{\qT}$ \cite{Grmela2019}. Note that $\qqP$ becomes an odd variable (although being originally even) after the reduction, similarly as $\qqJ$ becomes even after the reduction to the Fourier law although being originally odd. Inserting \re{rel:W} into \re{dwdt-GK-GEN}, one can recognise a GK-like relationship
\begin{align}
    \pdt \qqw &= - \pdr \qT + \pdr \left( \frac{\qT}{\qk_{11}} \pdr \qqJ \right) + \qk \left( - \frac{\qqJ}{\qT} \right) .
\end{align}
We remark, that we obtained the entropy current density -- heat current density relationship \re{GK:q-JJ}, which is equivalent to \re{GK:q-J} or \re{j(q,C)} and, entropy current density $ \qqJJ $ purely the conjugate to $ \qqw $ is.

If fast evolution of $\qqw$ is also already relaxed then the Ginzburg--Landau type heat conduction can be recognised, \viz
\begin{align}
    0 &= - \pdr \qT + \pdr \left( \frac{\qT}{\qk_{11}} \pdr \qqJ \right) + \qk \left( - \frac{\qqJ}{\qT} \right) .
\end{align}

In these approximations, $\qqP$ inherits the time dependence from $\qT$ and $\qqJ$ while $\qqP$ and $\qqJ$ inherit the time dependence from $\qT$, respectively. Thus the heat current density -- entropy current density relationship $ \qqq = \left( \qT \ident + \qqP \right) \qqJ $ is revealed. This means that, via reducing a more detailed description, a complete agreement can be found among the theories.

\subsubsection{Generalized ballistic--diffusive heat conduction} \label{BC-ext}

A  non-local generalisation of ballistic--diffusive heat conduction equation via using a third-order entropy current multiplier was derived in \cite{Kovacs2015}. Namely, a generalised entropy \re{s(e,q,Q)} and an extended entropy current density
\begin{align}
    \label{j(q,C1,C2)}
    \qqJJ = \left( \frac{1}{T} \ident + \qqC \right) \Cdot \qqq + \bbC \qodot \qqQ , 
\end{align}
with the third-order Ny\'\i ri multiplier $\bbC$, yield the entropy production density rate
\begin{align}
    \label{s-pr-balcond-ext}
    \qSig = \qqq \Cdot \left( -m_1 \pdt \qqq + \pdr \frac{1}{\qT} + \pdr \qqC \right) + \qqQ \Cdot \left( - m_2 \pdt \qqQ + \pdr \bbC \right) + \qqC \qodot \pdr \qqq + \bbC \qodot \pdr \qqQ ,
\end{align}
where the first term is a vector--vector coupling, the last one is a third-order--third-order one, and the rest terms prescribe second-order tensorial couplings. (Again, all these are three space dimensional terminologies.) Positive semi-definiteness of \re{s-pr-balcond-ext} is ensured by the equations
\begin{subequations}
\begin{align}
    \label{Ons-x_3-ext-1}
    - m_1 \pdt \qqq + \pdr \frac{1}{\qT} + \pdr \qqC &= \ql \qqq , \\
    \label{Ons-x_3-ext-2}
    - m_2 \pdt \qqQ + \pdr \bbC &= \qql{11} \qqQ + \qql{12} \pdr \qqq , \\
    \label{Ons-x_3-ext-3}
    \qqC &= \qql{21} \qqQ + \qql{22} \pdr \qqq , \\
    \label{Ons-x_3-ext-4}
    \bbC &= \qL \pdr \qqQ
\end{align}
\end{subequations}
with $ \qL \ge 0 $ and \re{req:l}. After eliminating entropy current multipliers $\qqC$ and $\bbC$,  we have
\begin{align}
    \label{d(q,Q)dt-ext}
    \begin{pmatrix} - m_1 \pdt \qqq \\ - m_2 \pdt \qqQ \end{pmatrix} =
    \begin{pmatrix} - \pdr \frac{1}{\qT} - \qla \pdr \qqQ \\ - \qla \pdr \qqq \end{pmatrix} +
    \begin{pmatrix} \ql \qqq - \qls \pdr \qqQ - \qql{22} \pdr^2 \qqq \\ \qql{11} \qqQ + \qls \pdr \qqq - \qL \pdr^2 \qqQ \end{pmatrix} .
\end{align}
Similarly to \re{d(q,Q)dt}, the first and second terms of RHS of \re{d(q,Q)dt-ext} represent the reversible and irreversible contributions to time evolution, respectively. Let the operator $\left( m_2 \pdt + \qql{11} - \qL \pdr^2 \right)$ act on the first equation of \re{d(q,Q)dt-ext}: then we obtain the constitutive equation on heat current density, \ie
\begin{align}
    \nonumber
    m_1 m_2 \pdt^2 \qqq + \left( m_1 \qql{11} + m_2 \ql \right) \pdt \qqq &- \left( m_1 \qL + m_2 \qql{22}\right) \pdt \pdr^2 \qqq - \left( \qql{11}\qql{22} - \qql{12}\qql{21} \right) \pdr^2 \qqq - \qql{22} \qL \pdr^4 \qqq + \ql \qql{11} \qqq \\
    &= m_2 \pdt \pdr \frac{1}{ \qT} + \qql{11} \pdr \frac{1}{\qT} - \qL \pdr^3 \frac{1}{\qT} .
\end{align}

We look for a non-local extension of the ballistic--diffusive heat conduction equation, consequently, total energy and the Hamiltonian structure of the reversible contribution presented in Sec. \ref{sec-BC} are not modified. On the other side, the dissipation potential is extended via a further non-local term, \ie
\begin{align}
    \qXi_{\rm eBD} \left( \qhat \qqxi_3^\qS{}^{^{^*}} \right) = \qXi_{\rm BD} \left( \qhat \qqxi_3^\qS{}^{^{^*}} \right) +  \int_\qL \frac{\qK}{2} \left[ \pdr \left( - \frac{\qqP}{\qT} \right) \right]^2 \dd \qr
\end{align}
with non-negative constant $\qK$. This dissipation potential leads to the time evolution equations
\begin{subequations}
\begin{align}
    \pdt \qeps &= - \pdr \left[ \left( \qT \ident + \qqP \right) \qqJ \right] , \\
    \pdt \qsig &= - \pdr \left\{ \qqJ + \left[ \qk_{12}^{\rm S}  \left( - \frac{\qqP}{\qT} \right) + \qk_{22} \pdr \left( - \frac{\qqJ}{\qT} \right) \right] \left( - \frac{\qqJ}{\qT} \right) + \qK  \pdr \left( - \frac{\qqP}{\qT} \right) \left( - \frac{\qqP}{\qT} \right) \right\} \\
    \nonumber
    & \hskip 5ex +  \qk \left( - \frac{\qqJ}{\qT} \right)^2 + \qk_{11} \left( - \frac{\qqP}{\qT} \right)^2 + 2 \qk_{12}^{\rm S} \left( - \frac{\qqP}{\qT} \right) \pdr \left( - \frac{\qqJ}{\qT} \right) + \qk_{22} \left[ \pdr \left( - \frac{\qqJ}{\qT} \right) \right]^2 + \qK \left[ \pdr \left( - \frac{\qqP}{\qT} \right) \right]^2 , \\
    \pdt \qqw &= - \pdr \qT - \pdr \qqP + \qk \left( - \frac{\qqJ}{\qT} \right) - \qk_{12}^{\rm S} \pdr \left( - \frac{\qqP}{\qT} \right) - \qk_{22} \pdr^2 \left( - \frac{\qqJ}{\qT} \right) , \\
    \label{dWdt-ext}
    \pdt \qqW &= - \pdr \qqJ + \qk_{11} \left( - \frac{\qqP}{\qT} \right) + \qk_{12}^{\rm S} \pdr \left( - \frac{\qqJ}{\qT} \right) - \qK \pdr^2 \left( - \frac{\qqP}{\qT} \right) .
\end{align}
\end{subequations}
Heat and entropy current densities
\begin{align}
    \qqq &= \left( \qT \ident + \qqP \right) \qqJ , &
    \qqJJ &= \qqJ + \left[ \qk_{12}^{\rm S}  \left( - \frac{\qqP}{\qT} \right) + \qk_{22} \pdr \left( - \frac{\qqJ}{\qT} \right) \right] \left( - \frac{\qqJ}{\qT} \right) + \qK  \pdr \left( - \frac{\qqP}{\qT} \right) \left( - \frac{\qqP}{\qT} \right)
\end{align}
can be recognized. The third order tensorial coupling in the dissipation potential gives additional contribution to entropy current density and, based on the conclusions of Sec. \ref{sec-BC} the equivalence with \re{j(q,C1,C2)} can be obtained. Furthermore, the equivalence of time evolution equations holds.

\subsection{Generalized ballistic-conductive heat conduction reduced from a four-dimensional state space}

Instead of further generalisations, we show that a four-dimensional state space,  spanned by variables
\begin{align}
    \qhat \qqxi_4^\qE = \begin{pmatrix} \qsig \\ \qqw \\ \qqW \\ \bbW \end{pmatrix},
\end{align}
results in a time evolution system whose variable with highest tensorial order is reduced to obtain the non-local extension of ballistic--diffusive heat conduction equation presented in Sec. \ref{BC-ext}, and that the heat current density -- entropy current density relationship obtained from NET-IV and GENERIC is also the same. Here, $\bbW$ denotes a third-order tensorial state variable (when in three space dimensions), the physical interpretation of which is not known. Now let us formulate this model within the GENERIC framework.

Supposing that total energy $ \qE = \int_{\qV} \qeps \left( \qqxi_4^\qE \right) \dd \qV$ does not contain non-local terms, the partial derivatives
\begin{align}
    &&
    \qT &= \pder{\qeps}{\qsig}{\qqw,\qqW,\bbW} , &
    \qqJ &= \pder{\qeps}{\qqw}{\qsig,\qqW,\bbW} , &
    \qqP &= \pder{\qeps}{\qqW}{\qsig,\qqw,\bbW} , &
    \bbP &= \pder{\qeps}{\bbW}{\qsig,\qqw,\qqW}
    &&
\end{align}
Assuming a canonical structure, we obtain the time evolution of $\qqxi_4^\qE$ restricted to reversible dynamics, \viz
\begin{align}
    \begin{pmatrix} \pdt \qsig |_{\rm rev} \\ \pdt \qqw |_{\rm rev} \\ \pdt \qqW |_{\rm rev} \\ \pdt \bbW |_{\rm rev} \end{pmatrix} = 
    \begin{pmatrix} 0 & -\pdr & 0 & 0 \\ -\pdr & 0 & - \pdr & 0 \\ 0 & -\pdr & 0 & - \pdr \\ 0 & 0 & - \pdr & 0 \end{pmatrix} \begin{pmatrix} T \\ \qqJ \\ \qqP \\ \bbP \end{pmatrix} .
\end{align}
This means, due to the Onsager-Casimir reciprocal relation that $\bbW$ is a quantity odd \wrt TRT.

Variables in entropic representation and entropy conjugated variables are
\begin{align}
    &&
    \qqxi_4^\qS &= \begin{pmatrix} \qeps \\ \qqw \\ \qqW \\ \bbW \end{pmatrix} , & \Longrightarrow &&
    \qqxi_4^\qS{}^{^{^*}} &= \var{\qS}{\qqxi_3^\qS} = \begin{pmatrix} \frac{1}{\qT} \\ - \frac{\qqJ}{\qT} \\ - \frac{\qqP}{\qT} \\ - \frac{\bbP}{\qT} \end{pmatrix} ,
    &&
\end{align}
respectively. The chosen dissipation potential
\begin{align}
    \qXi_{\rm 4D} \left( \qqxi_4^\qS{}^{^{^*}} \right) = \qXi_{\rm BD} \left( \qqxi_3^\qS{}^{^{^*}} \right) + \int_\qL \frac{\qK}{2} \left( - \frac{\bbP}{\qT} \right)^2 \dd \qr 
\end{align}
and the above-presented Hamiltonian structure provide the time evolution of $\qhat \qqxi_4^\qE$, \ie
\begin{align}
    \nonumber
    \pdt \qsig &= - \pdr \left\{ \qqJ + \left[ \qk_{12}^{\rm S}  \left( - \frac{\qqP}{\qT} \right) + \qk_{22} \pdr \left( - \frac{\qqJ}{\qT} \right) \right] \left( - \frac{\qqJ}{\qT} \right) \right\} \\
    & \hskip 5ex + \qk \left( - \frac{\qqJ}{\qT} \right)^2 + \qk_{11} \left( - \frac{\qqP}{\qT} \right)^2 + 2 \qk_{12}^{\rm S} \left( - \frac{\qqP}{\qT} \right) \pdr \left( - \frac{\qqJ}{\qT} \right) + \qk_{22} \left[ \pdr \left( - \frac{\qqJ}{\qT} \right) \right]^2 + \qK \left( - \frac{\bbP}{\qT} \right)^2 , \\
    \pdt \qqw &= - \pdr \qT - \pdr \qqP + \qk \left( - \frac{\qqJ}{\qT} \right) - \qk_{12}^{\rm S} \pdr \left( - \frac{\qqP}{\qT} \right) - \qk_{22} \pdr^2 \left( - \frac{\qqJ}{\qT} \right) , \\
    \label{dWdt-4D}
    \pdt \qqW &= - \pdr \qqJ - \pdr \bbP + \qk_{11} \left( - \frac{\qqP}{\qT} \right) + \qk_{12}^{\rm S} \pdr \left( - \frac{\qqJ}{\qT} \right) , \\
    \pdt \bbW &= - \pdr \qqP + \qK \left( - \frac{\bbP}{\qT} \right) .
\end{align}
Time evolution of internal energy density is
\begin{align}
    \pdt \qeps = - \pdr \left[ \left( \qT \ident + \qqP \right) \qqJ + \bbP \qqP \right] ,
\end{align}
thus heat and entropy current densities are
\begin{align}
    \qqq &= \left( \qT \ident + \qqP \right) \qqJ + \bbP \qqP , &
    \qqJJ &= \qqJ + \left[ \qk_{12}^{\rm S}  \left( - \frac{\qqP}{\qT} \right) + \qk_{22} \pdr \left( - \frac{\qqJ}{\qT} \right) \right] \left( - \frac{\qqJ}{\qT} \right) .
\end{align}
As we have expected, no additional terms appear in the entropy current density with respect to the ballistic--diffusive one, but a third-order tensorial contribution appears in the heat current density.

If fast evolution of $\bbW$ is relaxed then
\begin{align}
    0 = - \pdr \qqP + \qK \left( - \frac{\bbP}{\qT} \right) ,
\end{align}
which is the analogue of \re{Ons-x_3-ext-4} and can be treated as a constitutive relation among $\qqP$ and $\frac{\bbP}{\qT}$. Accordingly, substituting $\bbP = - \frac{\qT}{\qK} \pdr \qqP $ into \re{dWdt-4D}, we obtain
\begin{align}
    \pdt \qqW &= - \pdr \qqJ + \qk_{11} \left( - \frac{\qqP}{\qT} \right) + \qk_{12}^{\rm S} \pdr \left( - \frac{\qqJ}{\qT} \right) + \pdr \left( \frac{\qT}{\qK} \pdr \qqP \right) ,
\end{align}
which has the same structure as the second equation of \re{d(q,Q)dt-ext}, and with the above approximations heat current density -- entropy current density relationships obtained in both frameworks proved to be equivalent.

GENERIC allows for construction of a hierarchy of models of heat conduction varying in the level of detail involved, and the more detailed models (with more state variables) can be reduced to the less detailed. This is a demonstration of the multiscale nature of GENERIC.

\section{Conclusions}

In this paper, first, we revisited known derivations of beyond-Fourier heat conduction in the frameworks of NET-IV and of GENERIC, and we analysed their similarities and differences. 

Similarly to CIT, NET-IV determines the entropy production density rate from a concave expression of entropy and an assumed entropy current density -- heat current density relationship constraint, utilizing the classical balances on mass, linear and angular momenta, and internal energy, \eg in case of the Navier--Stokes--Fourier system and its extensions, like in \cite{Kovacs2021}.
Since entropy is expressed as a concave function of the state variables, the chosen representation is usually the entropic representation. Positive semi-definiteness of the derived entropy production rate density is ensured by Onsagerian equations. When internal variables, for which no balance equation is known, are also variables of entropy, then the Onsagerian equations contain time derivative terms, too. As we have shown,
when no assumption is made regarding reciprocal relations on the coefficients of the Onsagerian equations, then
the methodology of NET-IV generates both entropy preserving and entropy increasing terms in the time evolution equations (see, in this respect, \cite{Van2020}). Very often, internal variables have no known physical interpretation, hence, measuring them and performing calculations with them (without the knowledge of the related coefficients) is not feasible. Assuming a linear relation among the internal variable and its conjugate---which is almost always built in the concave expression of entropy---, internal variables and entropy current multipliers can be eliminated, thus extended constitutive laws can be obtained for the usual, known, and measurable physical quantities.

In contrast to NET-IV, GENERIC yields all evolution equations from the two generators, total energy and entropy (and the related operator matrices), from a Poisson bracket (or bivector) and a dissipation potential. GENERIC gives the non-dissipative, entropy-preserving, part and the irreversible, entropy-increasing, part of time evolution separately; the first one is convenient in the variables of the energetic representation while the second one in the entropic representation. The Hamiltonian structure of the reversible contribution is expressed by Poisson brackets, which can often be deduced from microscopic theories or by means of differential geometry on Lie groups.
The irreversible contribution to time evolution can be deduced from a usually
convex dissipation potential, which may also be determined from statistical theories \cite{Mielke2014}. In case of a quadratic dissipation potential, the generated system of time evolution is linear in the conjugate variables.

The differences between the two approaches are summarised in Table \ref{tab:dif}.
\vskip -2ex
\begin{table}[H]
    \centering
	\begin{tabular}{c||c|c}
		&NET-IV&GENERIC\\
		\hline
		\hline
		Representation & Entropy representation & Both energy and entropic representation \\
		\hline
		\multirow{3}{*}{Input}& Internal energy entropy relationships & Internal energy entropy relationships \\
		& Extended relationship on $\qqJ$ & Poisson bracket \\
		& Classical balance equations & Dissipation potential \\
		\hline
		\multirow{2}{*}{Output}& Entropy production rate density & Time evolution equations \\
		& Time evolution equations & Relationship of $\qqJ$ and $\qqq$ \\
		\hline
	\end{tabular}
	\caption{Differences between the NET-IV and GENERIC approaches. $\qqJ$: entropy current density, $\qqq$: heat current density.}
    \label{tab:dif}
\end{table}

The two derivations have led to the same structure of evolution equations and to the same relationship between entropy current density and heat current density.

Reassuringly, wherever we have applied approximation or reduction, thermodynamical consistency has been preserved.

GENERIC recognises both heat and entropy current densities from the evolution equations of internal energy and entropy densities and, connects them to the conjugate of the vectorial variable $ \qqw $. After algebraic reformulations the entropy current density -- heat current density relationship can be obtained. This derivation has lead us to our main recognition. 

When the dissipation potential contains nonlocal terms, which has higher tensorial orders than the state variable with highest tensorial order, then nonlocal extensions will appear in the generated evolution equations, as well as in the relationship between entropy current density and heat current density also higher order multipliers can be obtained, see \eg. MCV and GK equations for two dimensional state space and BC and its generalisation for three dimensional state space. To get the same relations with algebraic dissipation potentials, we had to extend the state space with a variable of tensorial order larger than the highest tensorial order of the variables of the starting state space. Subsequently, deriving an evolution equation for this variable, and reducing it, the resulting entropy current density -- heat current density relationships are also found equivalent, see \eg GK equation reduced from BC heat conduction.

After the fast attenuation of the variable with the highest tensorial order, its time dependence becomes enslaved by the variables with lower tensorial orders, thus explicit time dependence of this variable disappears, and the variable typically changes its parity \wrt TRT. However, the implicit time dependence of that variable contributes to the entropy current density -- heat current density relationship. In this limit, a spatial partial differential equation is obtained, which can be treated as a constitutive equation among the quantities which are included in this equation. Assuming a generalized entropy current density within NET-IV, only the spatial contributions of its variables appear in the entropy production rate density. Based on these, \textit{entropy current multipliers can be interpreted as relaxed state variables}.

Heat current density is frequently identified with a vectorial internal variable, however, we have shown that, from a general theoretical point of view, \textit{identification of heat current density with the conjugate of the vectorial internal variable is more advantageous}.

Since known GENERIC realizations of beyond-Fourier heat conduction treat models up to the GK equation \cite{Lebon2017,Grmela2019}, we formulated also the \textit{ballistic--diffusive heat conduction model in GENERIC}.

The present work strengthens the GENERIC background for beyond-Fourier heat conduction models, with a possible (and hoped) consequence that microscopic/multiscale (see, \eg \cite{Ottinger2005,Grmela2019, Pavelka2018, Klika2019}) understanding of these phenomena becomes available. In parallel, 
the GENERIC formulation opens a way towards new and efficient numerical methods (for example, along the lines of \cite{Portillo2017,Betsch2019}) for heat-conduction problems (like in \cite{Balassa2021}).

\section*{Nomenclature}

\begin{minipage}[t]{0.1\textwidth}
	$ \pdt , \ \pdr$ \\
	$ \ident $ \\
	$ \qqC $ \\
	$ \bbC $ \\
	$ \qqJ $ \\
	$ \qqJJ $ \\
	$ e $ \\
	$ E $ \\
	$ k , \ K , \ l , \ m $ \\
	$ \qqL $ \\
	$ \qqM $ \\
	$ \qqP $ \\
	$ \bbP $ \\
	$ \qqq $ \\
	$ \qqQ $ \\
	$ s $ \\
    $ S $ \\
    $ ^{\rm S} , \ ^{\rm A} $ \\
    \\
    $ T $ \\
\end{minipage}
    \hfill
\begin{minipage}[t]{0.38\textwidth}
	partial time/space derivative \\
	identity tensor \\
	entropy current multiplier, second order \\
	entropy current multiplier, third order \\
	energy conjugated variable to $ \qqw $ \\
	entropy current density \\
	specific internal energy \\
	energy functional \\
	coefficients \\
	Poisson bivector \\
	dissipative matrix \\
	energy conjugated variable to $ \qqW $ \\
	energy conjugated variable to $ \bbW $ \\
	heat current density \\
	second-order tensorial internal variable \\
	specific entropy \\
	entropy functional \\
	symmetric/antisymmetric part of a tensor \\
	temperature
\end{minipage}
    \hfill\hfill\hfill
\begin{minipage}[t]{0.1\textwidth}
    $ \qqw $ \\
    \\
    $ \qqW $ \\
    $ \bbW $ \\
    $ \qqx_n^\qS $ \\
    \\
    $ \qqy $ \\
    \\
    $ \qqz $ \\
    $ \strev{\qqx} , \ \stirr{\qqx} $ \\
    \\
    $ \frac{\delta}{\delta x} $ \\
    $ \varepsilon $ \\
    $ \qqeta $ \\
	$ \qqxi_n^\qE , \ \qqxi_n^\qS $ \\
	\\
    $ \qXi $ \\
    $ \qrho $ \\
    $ \sigma $ \\ 
    $ \Sigma $ 
\end{minipage}
    \hfill
\begin{minipage}[t]{0.38\textwidth}
    vectorial state variable, specific phonon momentum \\
    second-order tensorial state variable \\
    third-order tensorial state variable \\
	collection of $ n $ specific extensive state variables in the entropic representation \\
	specific extensive vectorial internal variable \\
	energy conjugated internal variable to $ \qqy $ \\
	reversible/irreversible contribution to dynamics \\
	functional derivative w.r.t. $ x $ \\
	internal energy density \\
	density corresponds to $ \qqy $ \\
	collection of $ n $ state variable densities in the energetic/entropic representation \\
	dissipation potential \\
	(mass) density \\
	entropy density \\
	entropy production rate density
\end{minipage}

\section*{Acknowledgement}

The research reported in this paper and carried out at BME has been supported by the grants National Research, Development and Innovation Office - NKFIH FK 134277, and by the NRDI Fund (TKP2020 NC,Grant No. BME-NC) based on the charter of bolster issued by the NRDI Office under the auspices of the Ministry for Innovation and Technology.
This paper was supported by the J\'anos Bolyai Research Scholarship of the Hungarian Academy of Sciences.

MP was supported by Czech Science Foundation, project no. 20-22092S, and by Charles University Research program No. UNCE/SCI/023.

\bibliographystyle{unsrt}
\bibliography{library-szm-200714
.bib}

\end{document}